\documentclass[intlimits,twoside,a4paper]{article}

\usepackage[T2A]{fontenc}
\usepackage[cp1251]{inputenc}
\usepackage[eqsecnum]{cmpj3}

\issue{2026}{29}{3}{33602}
\doinumber{10.5488/CMP.29.33602}
\title[The effect of  homopolymer adding]%
{The effect of homopolymer adding on aggregation behavior in amphiphilic  triblock copolymers containing rigid blocks %
}
\author[X.-G. Han,  H. Zhang, Z. H. Sun, M. Y. Zhu]
{X.-G. Han\orcid{0000-0002-2724-0114}\footnote{Corresponding author: \email{xghan0@163.com}}, H. Zhang\orcid{0009-0006-2892-8021}, Z. H. Sun\orcid{0000-0001-9249-3276}, M. Y. Zhu\orcid{0009-0009-9035-6898}}
\address{
 School of Science, Inner Mongolia University of Science and Technology, Baotou 014010, China}

\date{Received 18 April 2026; revised 25 June 2026; accepted 11 July 2026; published 28 September 2026}
\begin{document}

\maketitle

\begin{abstract}
 

The effect of B homopolymer adding on the aggregation behavior of amphiphilic coil/rod/coil BAB triblock copolymers was studied using lattice self-consistent field theory. It depends on the length of the hydrophobic rod block and copolymer concentration. Compared with the solutions, at low concentrations, homopolymer addition is favorable to the emergence of cubic and large lamellar micelles. Although it enriches the structural behavior, the rearrangement is suppressed. As rod blocks increase, the rearrangement is related to the orientation-dependent diffusion and inverse diffusion processes. At relatively high concentrations, adding homopolymers promotes rearrangement. For long rod block case, the arrangements related to inverse laminarization and the cooperative growth emerges. At high concentrations, for short rod block system, homopolymer addition is advantageous to micelle rearrangement concerned with order/disorder transition, and the rearrangements in the system of intermediate length and long rod blocks are suppressed. This work aids in understanding of the growth mechanism of micelles with rigid block cores. 
  %
\printkeywords
%
\end{abstract}

\section{Introduction}
Block copolymers have attracted a considerable attention due to their capability to self-assemble into various periodic nanostructures. In recent years, coil/rod block copolymers where the rod blocks are primarily derived from $\pi$ conjugated main chains, helical secondary structures, or aromatic groups, have been of particular interest. These materials exhibit a geometric asymmetry between rod and coil blocks as well as orientational interactions between anisotropic rod blocks. This leads to the phase behavior distinct from that of coil/coil systems \cite{Olsen2008}. Previous experimental and theoretical studies have observed interesting phases in rod/coil block copolymer melts, including smectic, zigzag, wavy, arrow, perforated lamellar, stripe, and bicontinuous cubic phases, alongside conventional phases such as lamellae, spheres, and cylinders~\cite{Jenekhe1998,Cai2017}. These nanostructures have potential applications in optoelectronics, photonic devices~\cite{Coakley2004,Segalman2009,Wang2018a,Zhang2018}, and drug delivery \cite{Blanazs2009,Elsabahy2012,Hamley2005,Lau2008}.

Adding homopolymers to block copolymer systems is an effective method for controlling morphology~\cite{Cao2024,Cai2024,Zhang2026}. Song et al. used a hybrid self-consistent field theory in rod/coil diblock copolymers and found that both coil and rod homopolymers could enhance the stability of ordered phases~\cite{Song2011}. Wu et al. studied rod/coil diblock copolymers mixed with rod or coil homopolymers using dissipative particle dynamics simulations, and found that simply changing the homopolymer length can effectively control microdomain lamellar spacing and thickness~\cite{Wu2014, Hayashi2019}. Other important studies focused on hierarchical liquid crystalline structures~\cite{Zhu2013,Zhang2015}. Homopolymers can not only improve the stability and change the symmetry of ordered phases, but also affect the equilibrium dynamics, including the micelle-unimer equilibrium and micelle fusion. Garcia et al. studied homopolymer-mediated diblock copolymer micelle-micelle interactions arising from homopolymer depletion, proposing that the ``fluffy'' nature of the micelle corona weakens depletion-induced destabilization~\cite{Garcia2020}. It has also been found that homopolymer adding can change the critical concentration for micelle fusion in diblock copolymers~\cite{Repollet-Pedrosa2013}. However, Yang et al. found that the effect of homopolymer depletion differs between diblock and triblock copolymers. The size of triblock copolymer micelles is not considerably influenced by homopolymer depletion~\cite{Yang2021}. To date, the aggregation behavior in blends of rod/coil amphiphilic triblock copolymers and homopolymers has not been reported.

Lattice self-consistent field theory (LSCFT) provides a quantitative framework for studying the self-assembly behavior of complex block copolymer systems~\cite{Chen2006, Han2010,Han2015, Han2020}. This method via transfer matrices simply describes the connection and stiffness of chains. Compared to  continuum theories~\cite{Wang2023,Wang2026}, LSCFT effectively overcomes limitations in characterizing complex topological structures of rod/coil block copolymers while maintaining computational efficiency, especially to the solution system~\cite{Han2022}. Using LSCFT, our research group discovered multiple rearrangement behaviors of micelles and proposed an underlying mechanism in the solutions of amphiphilic coil/rod/coil triblock copolymers~\cite{Han2022}. This mechanism is related to interfacial properties from the conformational asymmetry between rod and coil blocks and the fragmentation/fusion behavior. In this work, we study the  phase behavior of the blends containing rod/coil block copolymers and coil homopolymers. The above studies of depletion of homopolymer in flexible polymers took into account the case of long homopolymer chain~\cite{Garcia2020,Repollet-Pedrosa2013, Yang2021}. { Long homopolymer evidently affects the stability of micelles~\cite{Garcia2020}. Based on the character of multiple
	rearrangement of rod/coil block copolymer solutions, the effect of short homopolymer on the equilibrium
	aggregation behavior is studied which corresponds to the wet-brush regime~\cite{Xie2024,Hayashi2019}}.  We analyze the effect of adding B homopolymer on the aggregation behavior in amphiphilic coil/rod/coil BAB triblock copolymers, and found that the depletion of homopolymers depends on copolymer concentrations and hydrophobic rod block lengths. A non-monotonous  concentration-dependent effect of the addition of homopolymers and  the new mechanism of rearrangements are described.

\section{Theory} \label{sec2}
This section provides a brief description of LSCFT for the blending system of B homopolymer and amphiphilic coil/rod/coil BAB  symmetric triblock copolymers under the incompressible assumption. The system consists of ${{n}_{b}}$ copolymer chains, each with a degree of polymerization ${{N}_{b}}$. Each copolymer is composed of ${{N}_{\text{st}}}$ sticker monomers forming the A middle hydrophobic block and ${{N}_{\text{ns}}}$ non-sticker monomers forming each B end hydrophilic block, i.e., $N_b={N}_{\text{st}}+2{N}_{\text{ns}}$. Additionally, the system contains ${{n}_{h}}$ homopolymer chains, whose degree of polymerization is ${{N}_{h}}$. The total number of lattice sites is: ${{N}_{L}}={{n}_{b}}N_b+{{n}_{h}}{{N}_{h}}$.

The only non-bonded interaction is an attractive interaction $-\epsilon$ with $\epsilon>0$ between two nearest neighboring hydrophobic monomers in the system. An approximation is introduced for this attractive interaction~\cite{Han2010,Han2015}. Its expression is:
\begin{equation}
    \frac{U}{{{k}_{\text{B}}}T}=-\chi \sum\limits_{r}{{{\widehat{\phi }}_{st}}\left( r \right)}{{\widehat{\phi }}_{st}}\left( r \right),
\end{equation}
where $\chi =\frac{z}{2{{k}_{\text{B}}}T}\epsilon$, is the Flory-Huggins interaction parameter, $z$ is the lattice coordination number, $\sum_{r}$ denotes summation over the coordinates of all the lattice sites $r$, and ${{\widehat{\phi }}_{st}}\left( r \right)=\sum\nolimits_{j}{\sum\nolimits_{s\in st}{{{\delta }_{r,{{r}_{j,s}}}}}}$ is the volume fraction of hydrophobic segments on site $r$, where $s$ and $j$ and are the indexes of the segment and chain of a copolymer, respectively. Note that $s\in st$ means that the $s$-th segment is a sticker segment.

In this LSCFT, the copolymer is expressed as a random walk that forbids a direct backfolding, governed by a transfer matrix $\lambda $ which depends solely on the chain model. For a flexible subchain,
\begin{equation}
    \lambda _{{{r}_{j,s}}-r_{j,s-1}^{\prime}}^{{{\alpha }_{j,s}}-\alpha _{j,s-1}^{\prime}}=\left\{ \begin{matrix}
   0, & {{\alpha }_{j,s}}=\alpha _{j,s-1}^{\prime}\,,  \\
   1/\left( z-1 \right), & \text{otherwise} . \\
\end{matrix} \right.
\end{equation}
And for a rigid subchain,
\begin{equation}
    \lambda _{{{r}_{j,s}}-r_{j,s-1}^{\prime}}^{{{\alpha }_{j,s}}-\alpha _{j,s-1}^{\prime}}=\left\{ \begin{matrix}
   1, & {{\alpha }_{j,s}}=\alpha _{j,s-1}^{\prime}\,, \\
   0, & \text{otherwise}.  \\
\end{matrix} \right.
\end{equation}
Here, ${{r}{'}}$ is nearest neighboring site  $r$. ${{\alpha }_{s}}$ and ${{r}_{s}}$ are the bond orientation and position of the $s$-th monomer of the chain, respectively. While the chain self-intersections are forbidden, the effect of the excluded volume is appropriately introduced. Using the transfer matrix $\lambda$, the end monomer distribution functions of the $s$ th monomer of the polymer are calculated. For a copolymer and homopolymer, these are denoted as ${{G}^{{{\alpha }_{s}}}_{b}}\left( r,s | 1 \right)$ and ${{G}^{{{\alpha }_{s}}}_{h}}\left( r,s | 1 \right)$ , respectively.  They are obtained from the recursive relation:
\begin{equation}
    {{G}^{{{\alpha }_{s}}}}\left( r,s|1 \right)=G\left( r,s \right)\sum\limits_{r_{s-1}^{\prime}}{\sum\limits_{{{\alpha }_{s-1}}}{\lambda _{{{r}_{s}}-r_{s-1}^{\prime}}^{{{\alpha }_{s}}-{{\alpha }_{s-1}}}}}{{G}^{{{\alpha }_{s-1}}}}\left( {{r}{'}},s-1|1 \right),
\end{equation}
where $G\left( r,s \right)$ is the free monomer weighting factor,  differing for a copolymer and homopolymer. For a homopolymer $G_{h}\left( r,s \right)=\exp \left( -{{\omega }_{ns}}\left( {{r}_{s}} \right) \right)$. The expression  of a copolymer is
\begin{equation*}
    G_{b}\left( r,s \right)=\left\{ \begin{matrix}
   \exp \left( -{{\omega }_{st}}\left( {{r}_{s}} \right) \right), & s\in st,  \\
   \exp \left( -{{\omega }_{ns}}\left( {{r}_{s}} \right) \right), & s\in ns.  \\
\end{matrix} \right.
\end{equation*}
 The initial condition is ${{G}^{{{\alpha }_{1}}}}\left( r,1|1 \right)=G\left( r,1 \right)$ for all the ${{\alpha }_{1}}$. $\sum\nolimits_{r_{s-1}^{\prime}}{\sum\nolimits_{{{\alpha }_{s-1}}}}$ denotes summation over  all the possible positions and orientations of the $\left( s-1 \right)$-th monomer of the  polymer, respectively. 
 
 Another end monomer distribution function ${{G}^{{{\alpha }_{s}}}}\left( r,s|N \right)$, denoted by ${{G}_{b}^{{{\alpha }_{s}}}}\left( r,s|N_b \right)$ and ${{G}_{h}^{{{\alpha }_{s}}}}\left( r,s|N_{\text{h}} \right)$ for a copolymer and homopolymer, respectively, is calculated from:
 \begin{equation}
     {{G}^{{{\alpha }_{s}}}}\left( r,s|N \right)=G\left( r,s \right)\sum\limits_{r_{s+1}^{\prime}}{\sum\limits_{{{\alpha }_{s+1}}}{\lambda _{{{r}_{s+1}}-r_{s}^{\prime}}^{{{\alpha }_{s+1}}-{{\alpha }_{s}}}}}{{G}^{{{\alpha }_{s+1}}}}\left( {{r}{'}},s+1|N \right),
 \end{equation}
with the initial condition  ${{G}^{{{\alpha }_{N}}}}\left( r,N|N \right)=G\left( r,N \right)$ for all the ${{\alpha }_{N}}$.   The free monomer weighting factor $G\left( r,s \right)$ here is defined similar to the above one.

In the canonical ensemble, the free energy functional $F$ of the system is: 
\begin{equation}
    \frac{F\left[ {{\omega }_{+}},{{\omega }_{-}} \right]}{{{k}_{\text{B}}}T}=\sum\limits_{r}{\left\{ \frac{1}{4\chi }\omega _{-}^{2}\left( r \right)-{{\omega }_{+}}\left( r \right) \right\}}-{{n}_{b}}\ln {{Q}_{b}}\left[ {{\omega }_{st}},{{\omega }_{ns}} \right]-{{n}_{h}}\ln {{Q}_{h}}\left[ \omega _{ns} \right],
\end{equation}
where ${{Q}_{h}}$ is the single-chain partition function for a non-interacting homopolymer under the field ${{\omega }_{ns}}\left( r \right)={{\omega }_{+}}\left( r \right)$:
\begin{equation*}
     {{Q}_{h}}=\frac{1}{{{N}_{L}}}\frac{1}{z}\sum\limits_{{{r}_{N_{\text{h}}}}}{\sum\limits_{{{{{\alpha }_{N_{\text{h}}}}}}}{{{G}_h^{{{\alpha }_{N_{\text{h}}}}}}\left( r,N_{\text{h}}|1 \right)}},
\end{equation*}
${{Q}_{b}}$ is the single-chain partition function for a non-interacting copolymer chain subject to the fields ${{\omega }_{st}}\left( r \right)={{\omega }_{+}}\left( r \right)-{{\omega }_{-}}\left( r \right)$ and ${{\omega }_{ns}}\left( r \right)={{\omega }_{+}}\left( r \right)$, acting on the sticker and non-sticker segments, respectively:
\begin{equation*}
    {{Q}_{b}}=\frac{1}{{{N}_{L}}}\frac{1}{z}\sum\limits_{{{r}_{N_b}}}{\sum\limits_{{{r}_{{{\alpha }_{N_b}}}}}{{{G}^{{{\alpha }_{N}}}}\left( r,N_b|1 \right)}}.
\end{equation*}
Minimizing the expression of $F$ with respect to ${{\omega }_{-}}\left( r \right)$ and ${{\omega }_{+}}\left( r \right)$ yields the SCFT equations:
\begin{equation}
    {{\omega }_{-}}\left( r \right)=2\chi {{\phi }_{st}}\left( r \right),
    \label{eq2.7}
\end{equation}
\begin{equation}
    {{\phi }_{st}}\left( r \right)+{{\phi }_{ns}}\left( r \right)+\phi _{h}\left( r \right)=1,
     \label{eq2.8}
\end{equation}
where
\begin{equation}
    {{\phi }_{st}}\left( r \right)=\frac{1}{{{N}_{L}}}\frac{1}{z}\frac{{{n}_{b}}}{{{Q}_{b}}}\sum\limits_{s\in st}{\sum\limits_{{{\alpha }_{s}}}{\frac{{{G}_b^{{{\alpha }_{s}}}}\left( r,s|1 \right){{G}_b^{{{\alpha }_{s}}}}\left( r,s|N_b \right)}{G_b\left( r,s \right)}}},
\end{equation}
and
\begin{equation}
    {{\phi }_{ns}}\left( r \right)=\frac{1}{{{N}_{L}}}\frac{1}{z}\frac{{{n}_{b}}}{{{Q}_{b}}}\sum\limits_{s\in ns}{\sum\limits_{{{\alpha }_{s}}}{\frac{{{G}_b^{{{\alpha }_{s}}}}\left( r,s|1 \right){{G}_b^{{{\alpha }_{s}}}}\left( r,s|N_b \right)}{G_b\left( r,s \right)}}},
\end{equation}
and
\begin{equation}
    \phi _{h}\left( r \right)=\frac{1}{{{N}_{L}}}\frac{1}{{{z}}}\frac{{{n}_{h}}}{{{Q}_{h}}}\sum\limits_{1\leq s\leq{{N}_{h}}}{\sum\limits_{{{\alpha }_{s}}}{\frac{{{G}_h^{{{\alpha }_{s}}}}\left( r,s|1 \right){{G}_h^{{{\alpha }_{s}}}}\left( r,s|N_{\text{h}} \right)}{G_h\left( r,s \right)}}}.
\end{equation}
Here, ${{\phi }_{st}}\left( r \right)$,  ${{\phi }_{ns}}\left( r \right)$  and $\phi _{h}\left( r \right)$ represent the average volume fractions of  sticker monomers, non-sticker monomers, and B homopolymer monomers at site $r$, respectively.  

In this work, the SCFT equations are solved by using the real-space method on a cubic lattice with periodic boundary conditions. 
To identify the stable phase at any given point, SCFT computations are carried out using different initial potential
fields, which consist of the randomized potential fields and the fields from different saddle point configurations~\cite{Han2010, Han2015, Han2022}. The equilibrium phases obtained using all of these different initial conditions are examined for the free energy, and the one with the lowest free energy is identified as the stable phase.

Furthermore, two variables are introduced to characterize the aggregation behavior. The first variable is an order parameter~\cite{Dudo1999,Han2010}, which quantifies the aggregation extent of the system:
\begin{equation}
\Phi (\left\{ \phi _{st}(r)\right\} )=\frac{1}{N_{L}}\sum_{r}[\phi _{st}(r)-%
\bar{\phi}_{st}]^{2}=\frac{1}{N_{L}}\sum_{r}\phi _{st}^{2}(r)-\bar{\phi}%
_{st}^{2}.
\end{equation}%
Similar to the lattice model of living polymerization~\cite{Dudo1999},  the extent of aggregation  $\Phi $, defined as the fraction
of polymers converted into micelles, represents an order parameter type variable for associative system. The language is loose since there is no true phase transition associated with the associative behaviors.  Here, $\Phi $ is a function of the sticker volume fraction field $\left\{ \phi _{st}(r)\right\} $, which in turn depends on $\chi $ and $\bar{\phi}_{_{b}}$. For homogenous solutions, $\Phi $ equals zero.  As $\chi $ increases, $\Phi (\chi) $ should increase.  A decrease in $\Phi (\chi) $ signifies the reduction of  polymers converted into micelles, i.e., the  dissolution of  micelles. 

The second  is the heat capacity, which can signal structural transitions. Using equations (\ref{eq2.7}), (\ref{eq2.8}) and the expression of $\chi =\frac{z}{2{{k}_{\text{B}}}T}\epsilon $, the heat capacity per site (in units of ${{k}_{\text{B}}}$) for the coil/rod/coil triblock copolymer solution is defined as~\cite{Han2010,Han2015}:
\begin{equation}
  {{C}_{V}}=\frac{1}{{{N}_{L}}}{{\chi }^{2}}\frac{\partial }{\partial \chi }\left[ \sum\limits_{r}{\phi _{st}^{2}\left( r \right)} \right] . 
\end{equation}
  The inflection point of $\Phi $ versus $T$ can also be used to define the critical micelle temperature $T_c$ corresponding the maxinum of $C_V$. It is noticed that $\chi ^{2}\frac{\partial }{\partial {\chi }}$ is equivalent to $-\frac{\partial }{\partial (1/{\chi )}}$ and $\chi =\frac{z}{%
2k_{_\text{B}}T}\epsilon$. Therefore, $C_{_V}$ is proportional to the first derivative of $\Phi $ with respect to temperature, which
resembles the case of the living polymerization system~\cite{Dudo1999,Kennedy1983}. The formation of spherical micelles
from amphiphile molecules also has features in common with living polymerization~\cite{Dudo1999, Israelachvili1992}. Generally, the competition
between the energetic gain upon clustering and the entropic loss upon reducing the number of particles in the system is common to many equilibrium aggregating systems, including the living polymerization system, amphiphilic micelle formation and the clustering observed in
supercooled liquids~\cite{Kumar1999, Kumar2001,Han2010}. A peak in heat capacity indicates the changes in aggregate structures, particularly in micelle arrangement~\cite{Kumar2001,Han2010,Han2015}. Curves for ${{C}_{V}}$  and $\Phi $ will be calculated to characterize the micellar formation below for different chain architectures and polymer concentrations.
\section{Result and discussion} \label{sec3}
\begin{figure}[!t]
\centering
\includegraphics[width=7cm]{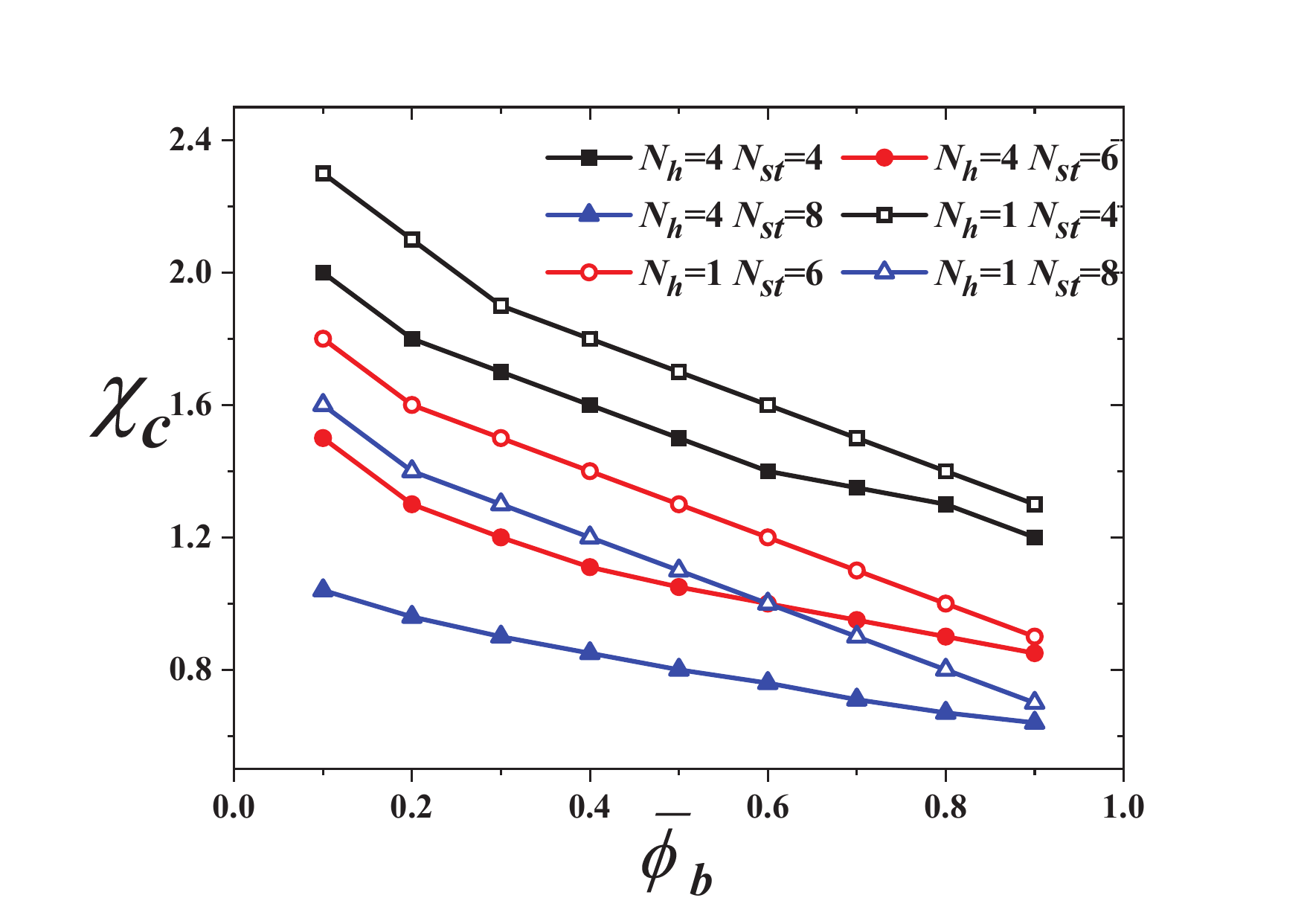}
\caption{(Colour online) Phase diagram of different hydrophobic rod block length $N_{\text{st}}$  blends of coil/rod/coil BAB triblock copolymers and homopolymers  compared with the corresponding triblock copolymer solutions.
\label{fig1}}
\end{figure}
This section discusses  the effect of homopolymer adding on the aggregation properties  in /coil/rod/coil BAB triblock copolymer melts compared to the  same triblock copolymer solutions. The system is governed by the following variables: the Flory-Huggins interaction parameter $\chi$, the copolymer length $N_b$ (held constant at $N_b=20$), the homopolymer length $N_{\text{h}}$ (held constant at  $N_{\text{h}}=4$), the middle hydrophobic block length $N_{\text{st}}=4,6,8$ and the concentration of
copolymers $\overline{\phi }_{b}=0.1, 0.3, 0.5, 0.8$. Simulations were performed on lattice sizes from $N_L = 40^3$ to $N_L = 60^3$ to ensure that the results were free from finite-size effects. As shown in figure~\ref{fig1}, the influence of homopolymer addition  depends on both the hydrophobic rod block length and the copolymer concentration. For a fixed rod block length, the effect of homopolymer addition is most pronounced at low and intermediate concentrations. Furthermore, the longer is the hydrophobic block, the more pronounced is the effect of homopolymer addition. Its effect is analogous to increasing copolymer concentration, leading to a decrease in the interaction parameter for unimer/micelle transition $\chi_c$.
\begin{figure}[!t]
\centering
\includegraphics[width=7cm]{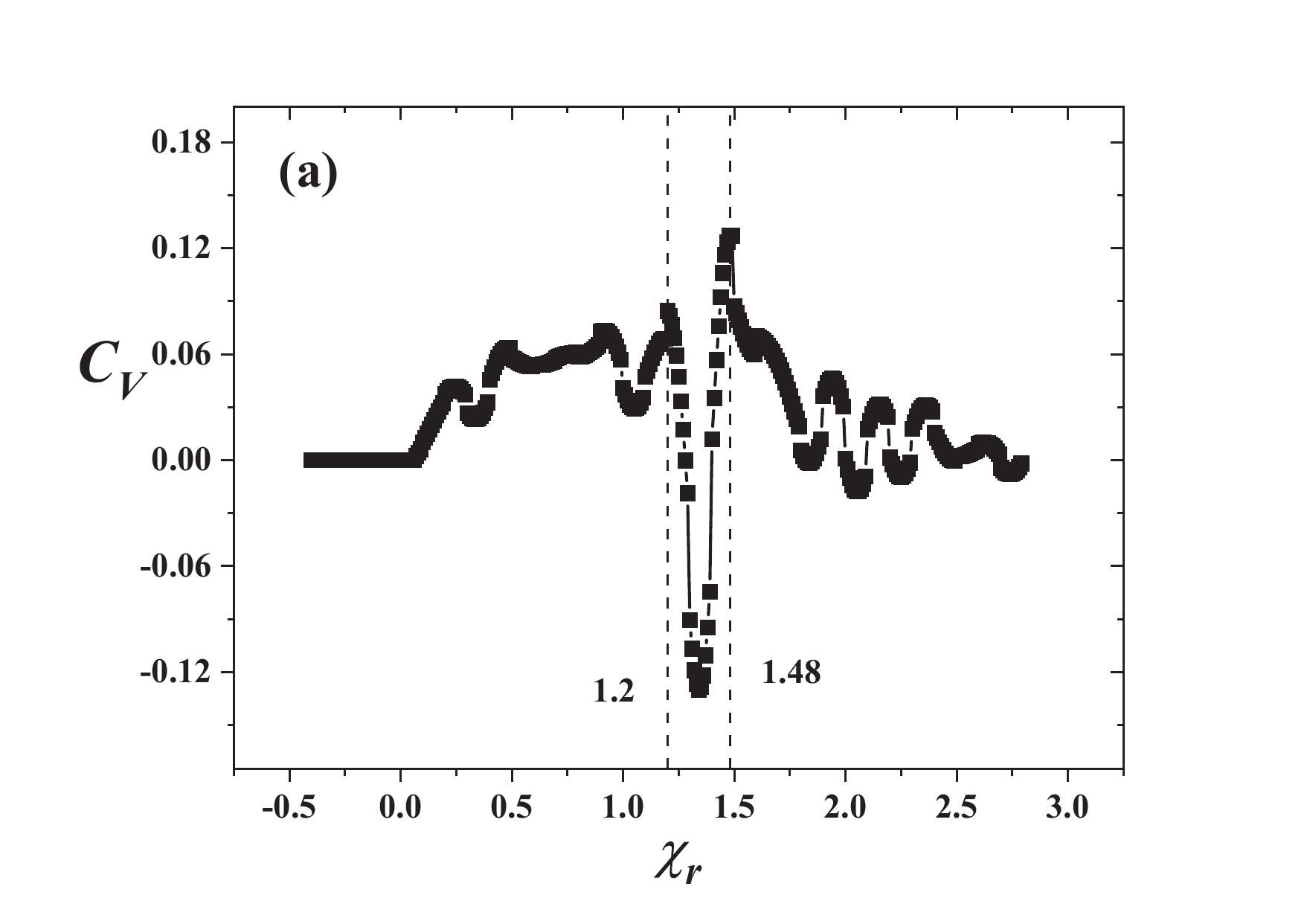}
\includegraphics[width=7cm]{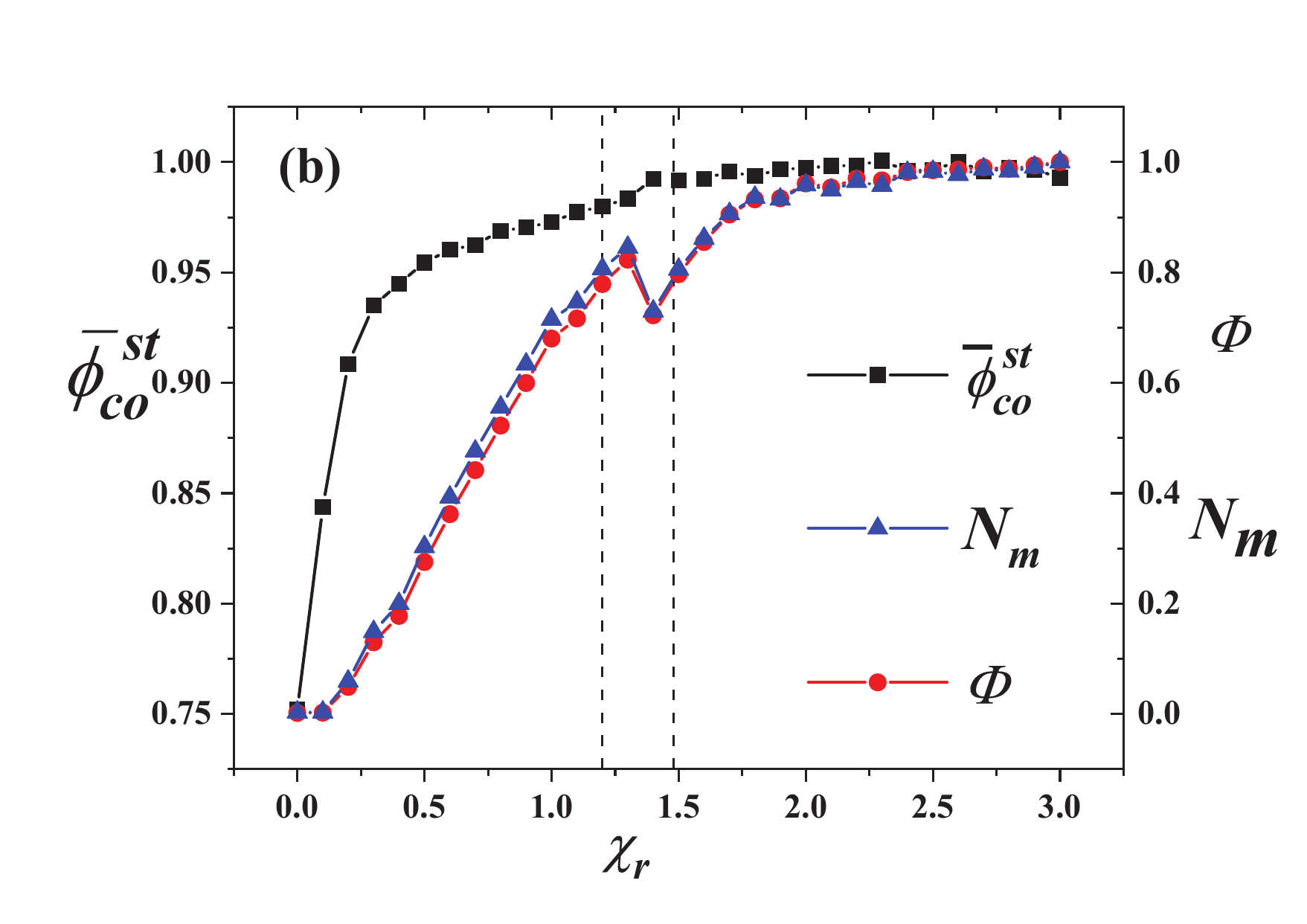}
\caption{(Colour online) Equilibrium aggregation curves for  $\overline{\phi }_{b}=0.1$ and ${{N}_{\text{st}}}=4$. (a): the  heat capacity curve $C_V$ of  $\chi_r$  ($\chi_r=\chi-\chi_c$). (b): the mean volume fraction of rod blocks in micellar cores $\overline{{{\phi }}}_{co}^{st}$,  effective volume of micellar cores ($N_m=N_m/{N^{\text{max}}_m}$) and order parameters change $\mathit{\Phi}$.
\label{fig2}}
\end{figure}

\begin{figure}[!t]
\centering
\includegraphics[width=12cm]{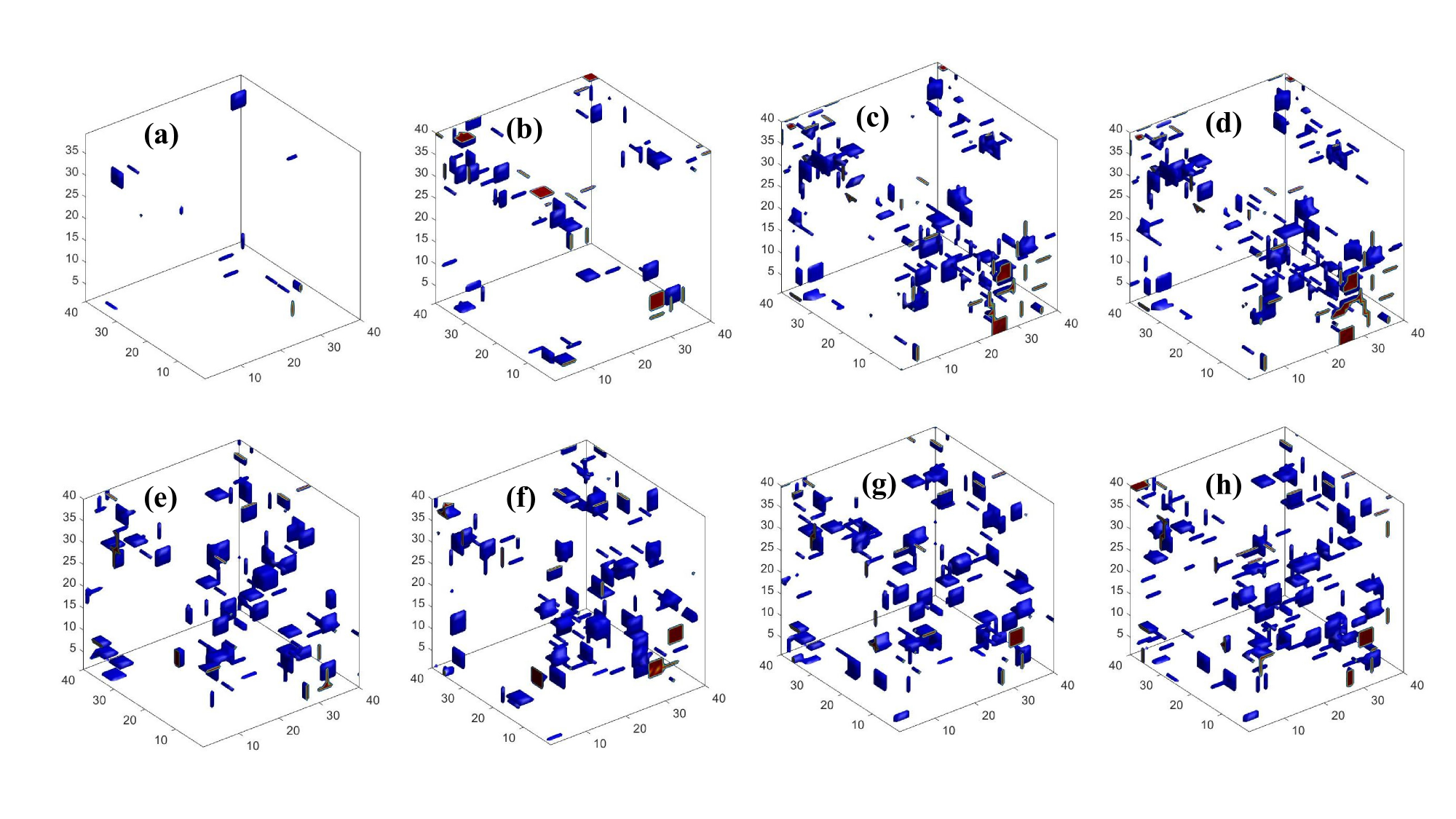}
\caption{(Colour online) The  micellar structures  for $\overline{\phi }_{b}=0.1$, ${{N}_{\text{st}}}=4$  (${{\phi }_{st}}>0.5$) under different interaction parameters. (a) ${{\chi }_{r}}=0.2$, (b) ${{\chi }_{r}}=0.6$, (c) ${{\chi }_{r}}=1.1$, (d) ${{\chi }_{r}}=1.3$, (e) ${{\chi }_{r}}=1.4$, (f) ${{\chi }_{r}}=1.6$, (g) ${{\chi }_{r}}=1.7$, (h) ${{\chi }_{r}}=2.2$.
\label{fig3}
}
\end{figure}

 At low concentrations ($\overline{\phi }_{b}=0.1$), for a short rod block ${{N}_{\text{st}}}=4$ (figure~\ref{fig2}a), the heat capacity curve shows three peaks (two positive, one negative), indicating multiple rearrangement events. When the interaction parameter ${{\chi }_{r}}$ increases, a few rods and lamellar micelles initially form, then grow, aggregate, and fuse within the first positive peak corresponding to (figure~\ref{fig3}a-c). At ${{\chi }_{r}}=1.4$, the rod micelles decrease while cubic micelles appear (figure~\ref{fig3}d,e), accompanied by the micelle fragmentation/fusion, as seen from the reduction in micellar core volume ${{N}_{m}}$ and  average aggregation degree $\overline{{{\phi }}}_{co}^{st}$ (figure~\ref{fig2}b). { While the decrease in aggregation degree $\Phi $  results in the negative heat capacity, the peak appearance corresponds to micelle dissolution.} Further micelle fusion produces the second positive peak (figure~\ref{fig3}f,g). Subsequent increases in ${{\chi }_{r}}$ lead to an increase in lamellar micelles, which break into rods or undergo the reverse process, corresponding to the remaining peaks (figure~\ref{fig3}h). Compared to the corresponding  solutions~\cite{Han2022}, the micelle diffusion behavior is largely absent, and transformations between rod and lamellar micelles are considerably weaker. It is noted that cubic micelles emerge at low concentrations. These properties are attributed to homopolymer depletion accelerating rod block aggregation, thereby weakening some rearrangement behavior~\cite{Miyazaki2022}.

\begin{figure}[!t]
\centering 
\includegraphics[width=5cm]{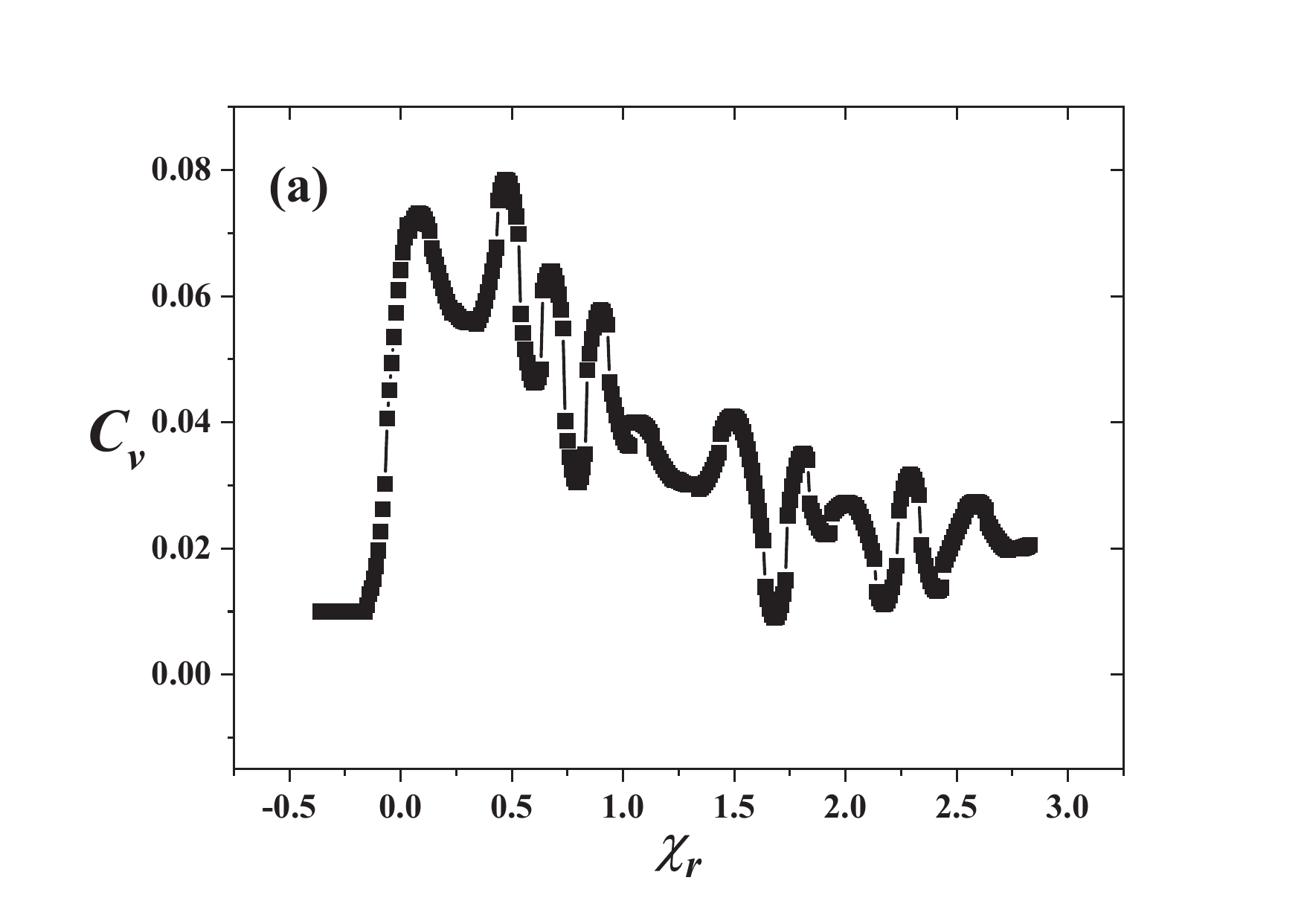}
\includegraphics[width=5cm]{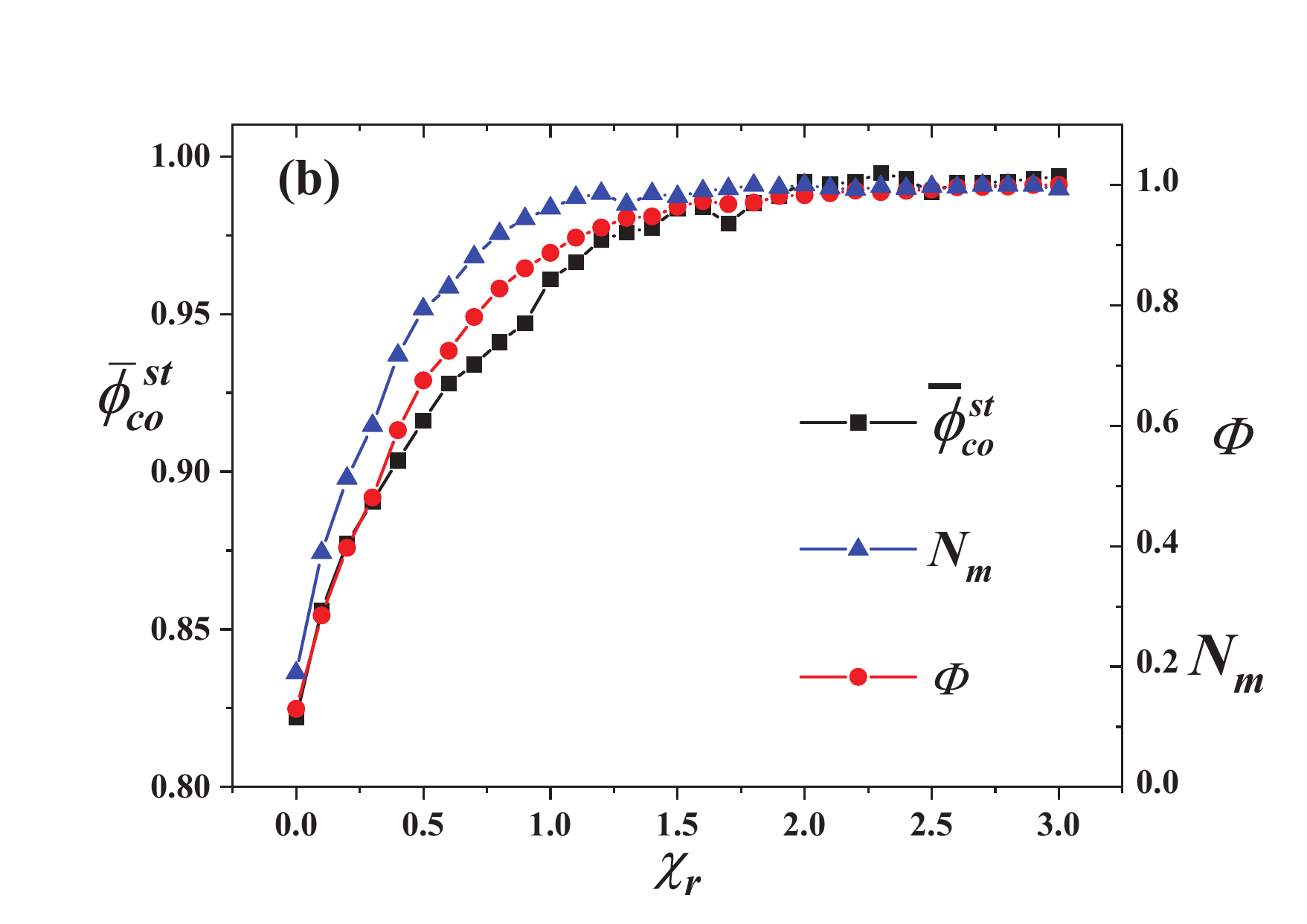}
\includegraphics[width=5cm]{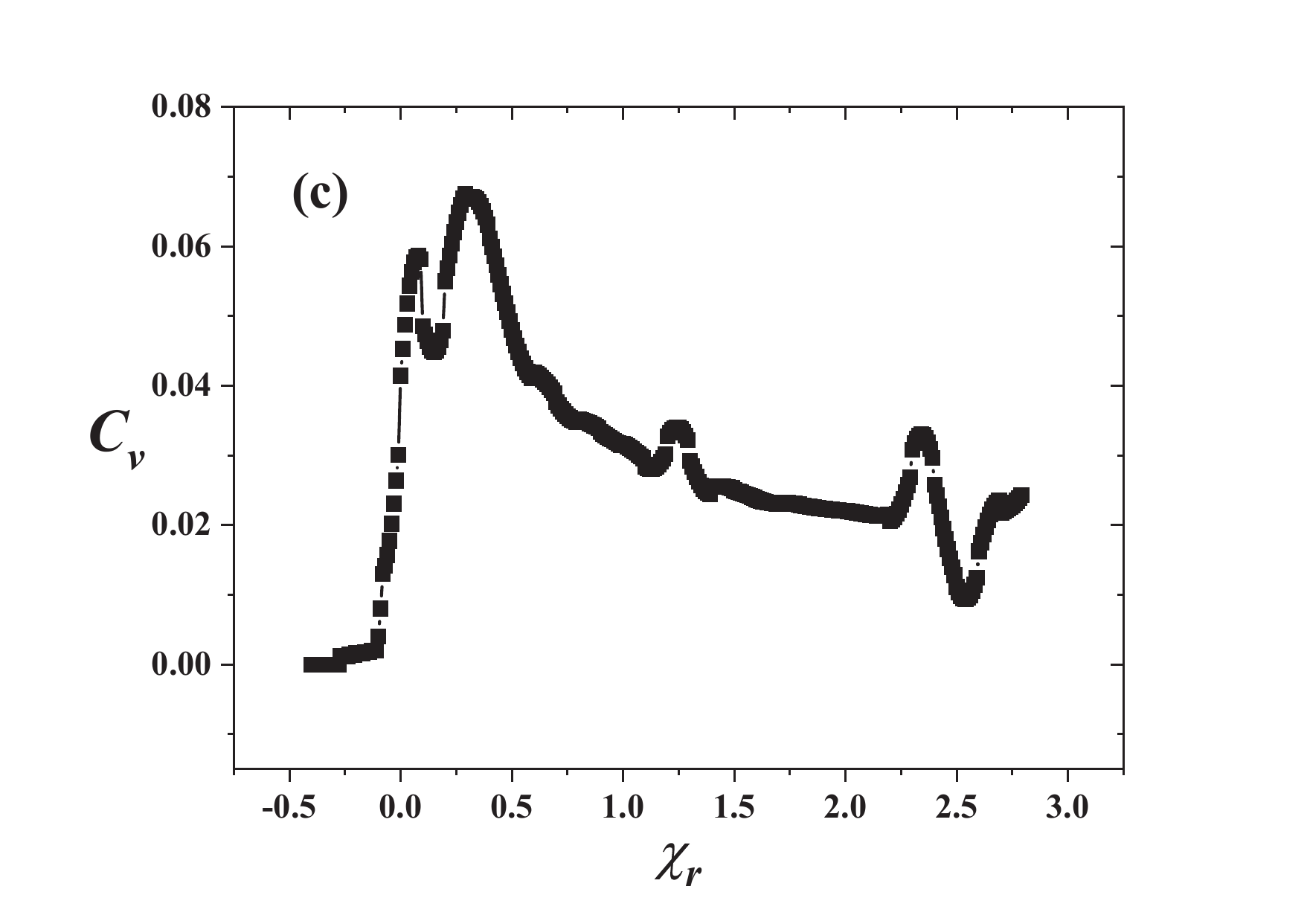}
\includegraphics[width=5cm]{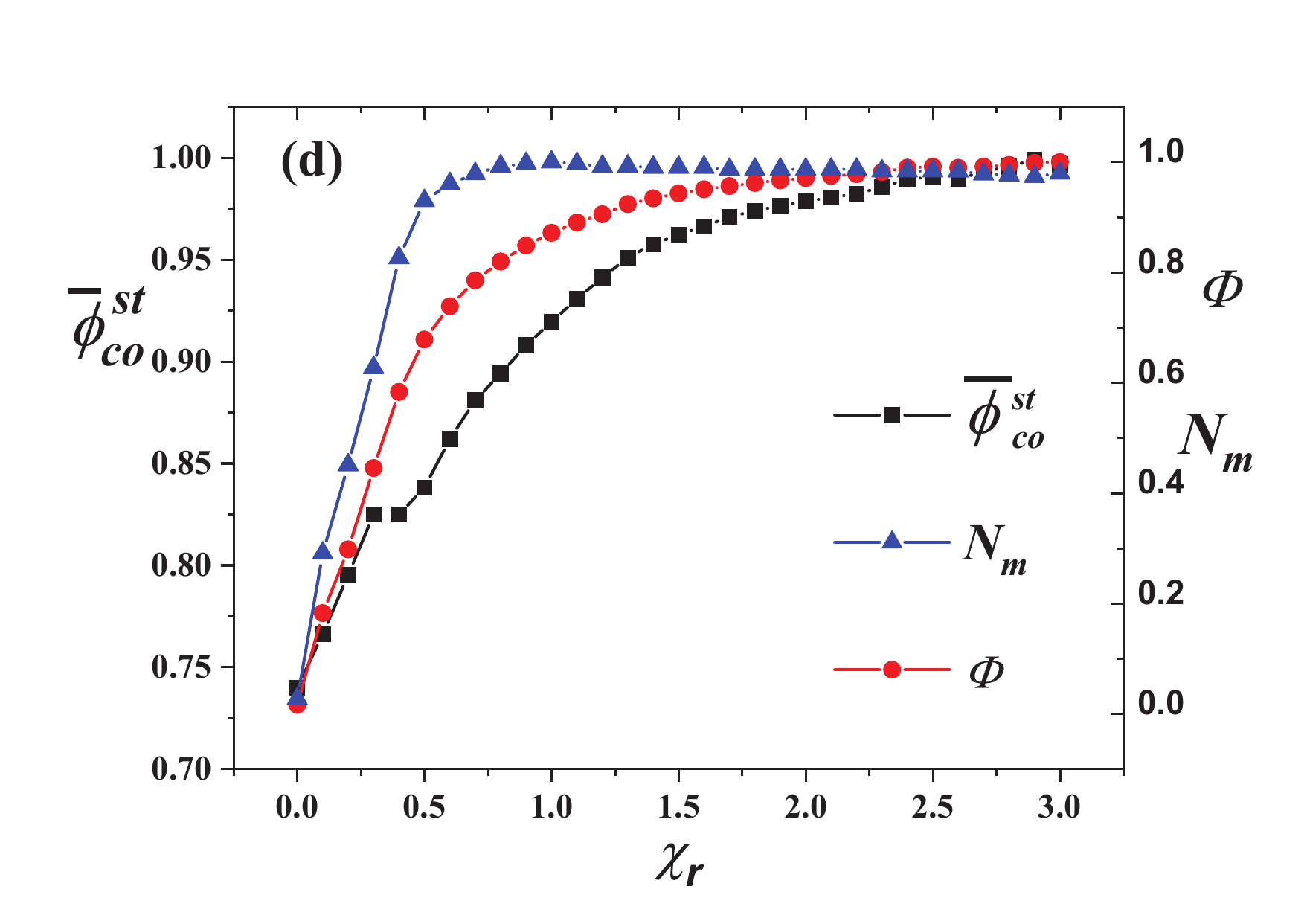}
\caption{(Colour online) Equilibrium aggregation curves of different ${{N}_{\text{st}}}$ for $\overline{{{\phi }}}_{b}=0.1$. (a) and (b) for ${{N}_{\text{st}}}=6$; (c) and (b) for ${{N}_{\text{st}}}=8$.} 
\label{fig4}
\end{figure}

\begin{figure}[!t]
\centering  
\includegraphics[width=12cm]{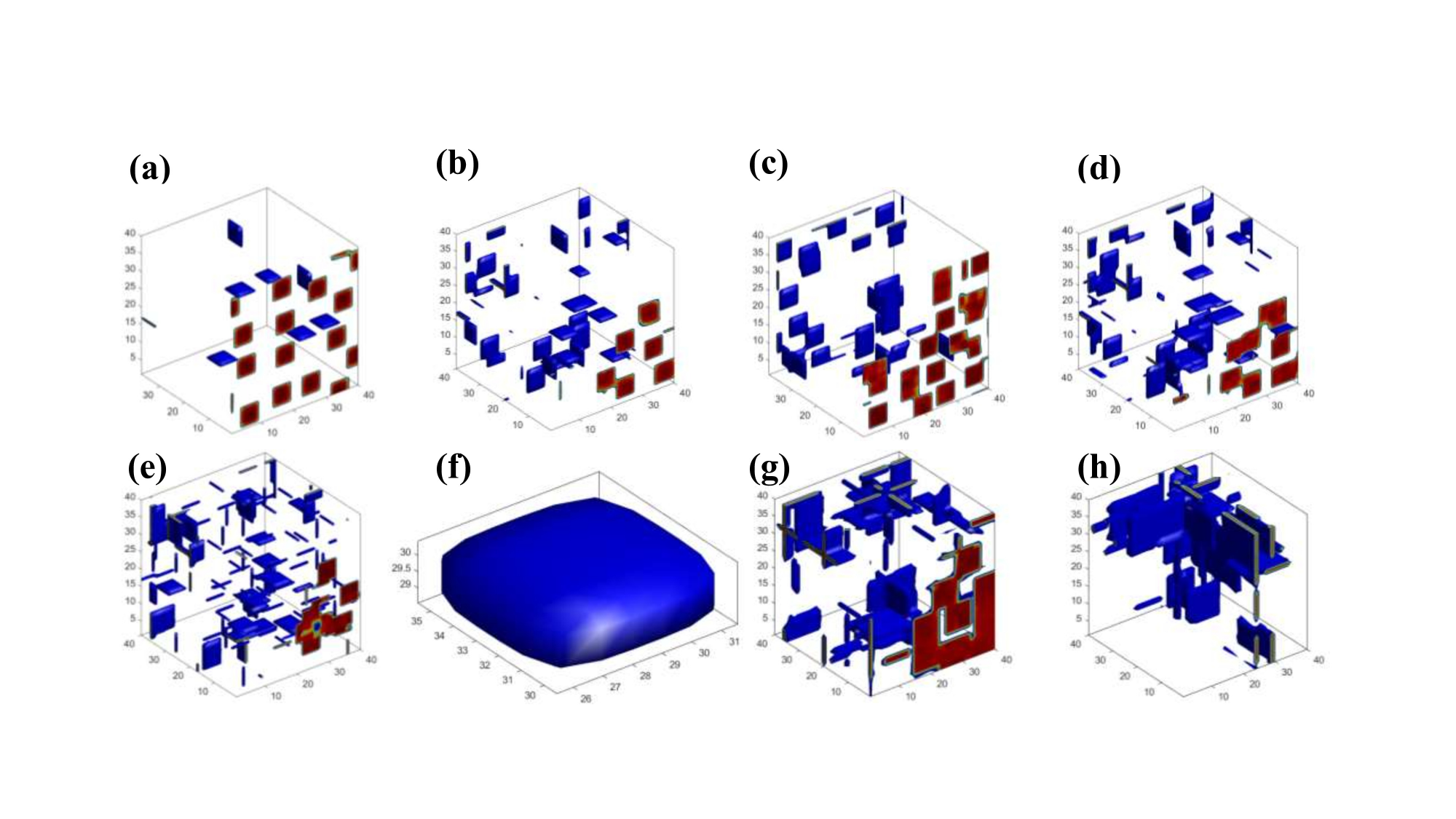}
\caption{(Colour online) The  micellar structures  for $\overline{\phi }_{b}=0.1$  (${{\phi }_{st}}>0.5$) under different ${{\chi }_{r}}$ and ${{N}_{\text{st}}}$. (a) ${{N}_{\text{st}}}=6$, ${{\chi }_{r}}=0.1$, (b) ${{N}_{\text{st}}}=6$, ${{\chi }_{r}}=0.3$, (c) ${{N}_{\text{st}}}=6$,  ${{\chi }_{r}}=0.5$, (d) ${{N}_{\text{st}}}=6$, ${{\chi }_{r}}=0.6$, (e) ${{N}_{\text{st}}}=6$, ${{\chi }_{r}}=1.4$,  (f) ${{N}_{\text{st}}}=8$, ${{\chi }_{r}}=0.0$, (g) ${{N}_{\text{st}}}=8$, ${{\chi }_{r}}=1.2$, (h) ${{N}_{\text{st}}}=8$, ${{\chi }_{r}}=1.3$.}
\label{fig5}
\end{figure}

For ${{N}_{\text{st}}}=6$ (figure~\ref{fig4}a), two  heat capacity peaks appear, where the disappearance of negative peaks indicates that the rearrangement weakens. The first peak has three sub-peaks: the first corresponds to the appearance of dispersed horizontal and vertical lamellar micelles (figure~\ref{fig5}a); the second corresponds to the aggregation, fusion, and compaction of these into a more vertically aligned state, i.e., a reverse diffusion process (figure~\ref{fig5}b,c); the third corresponds to a diffusion process (figure~\ref{fig5}d). The second peak has two sub-peaks associated with intense micelle fusion/fragmentation, leading to a considerable increase in rod micelles (figure~\ref{fig5}e). The remaining peaks relate to rod/lamella transformations, causing sudden fluctuations in the average aggregation degree (mean volume fraction of rod blocks in micellar cores), indicating the morphology changes driven by fragmentation/fusion (figure~\ref{fig4}b). { In the intermediate rod block length system, the homopolymer adding, which accelerates the aggregation of rod blocks, has the greatest effect on the lamellar micelles.} Therefore, the structural behaviors related to the change of lamellar orientation enrich the mechanism of the rearrangement behavior.


For long rod block (${{N}_{\text{st}}}=8$, figure~\ref{fig4}c), three peaks occur where sub-peaks basically disappear, which indicates that the aggregation behavior is less dependent on temperature. The first peak containing two sub-peaks is related to largely lamellar/lamellar and lamellar/rod morphology changes, accompanied by micellar fragmentation/fusion behavior, and changes in lamellar orientation (see figure~\ref{fig5}f--g). Within the  second peak, at ${{\chi }_{r}}=1.3$, it involves considerable aggregation and fusion of lamellar structures, and micelles occupy only part of the system space. { There is a reverse diffusion process that regularizes lamellar orientation (figure~\ref{fig5}h). At the same time, there is not a diffusion process with increasing $\chi_r$, which is different from that of the case with  ${{N}_{\text{st}}}=6$}. It is attributed to the acceleration of the aggregation of rod blocks, i.e., the emergence of larger lamellae. The third peak (${{\chi }_{r}}=1.61-2.4$), especially at ${{\chi }_{r}}=2.3$, shows no major structural change but a marked increase in aggregation degree, probably due to a further fragmentation/fusion (figure~\ref{fig4}d). Overall, rearrangement events decrease with the increasing rod length. { For sufficiently long rods, only the reverse diffusion process exists, indicating the system micellar growth is
	dominated by attractive rod-rod interactions.} In low concentrations, adding homopolymer leads to the emergence of the cubic and large lamellar micelles, and weakens the rearrangement.
\begin{figure}[!t]
\centering
\includegraphics[width=5cm]{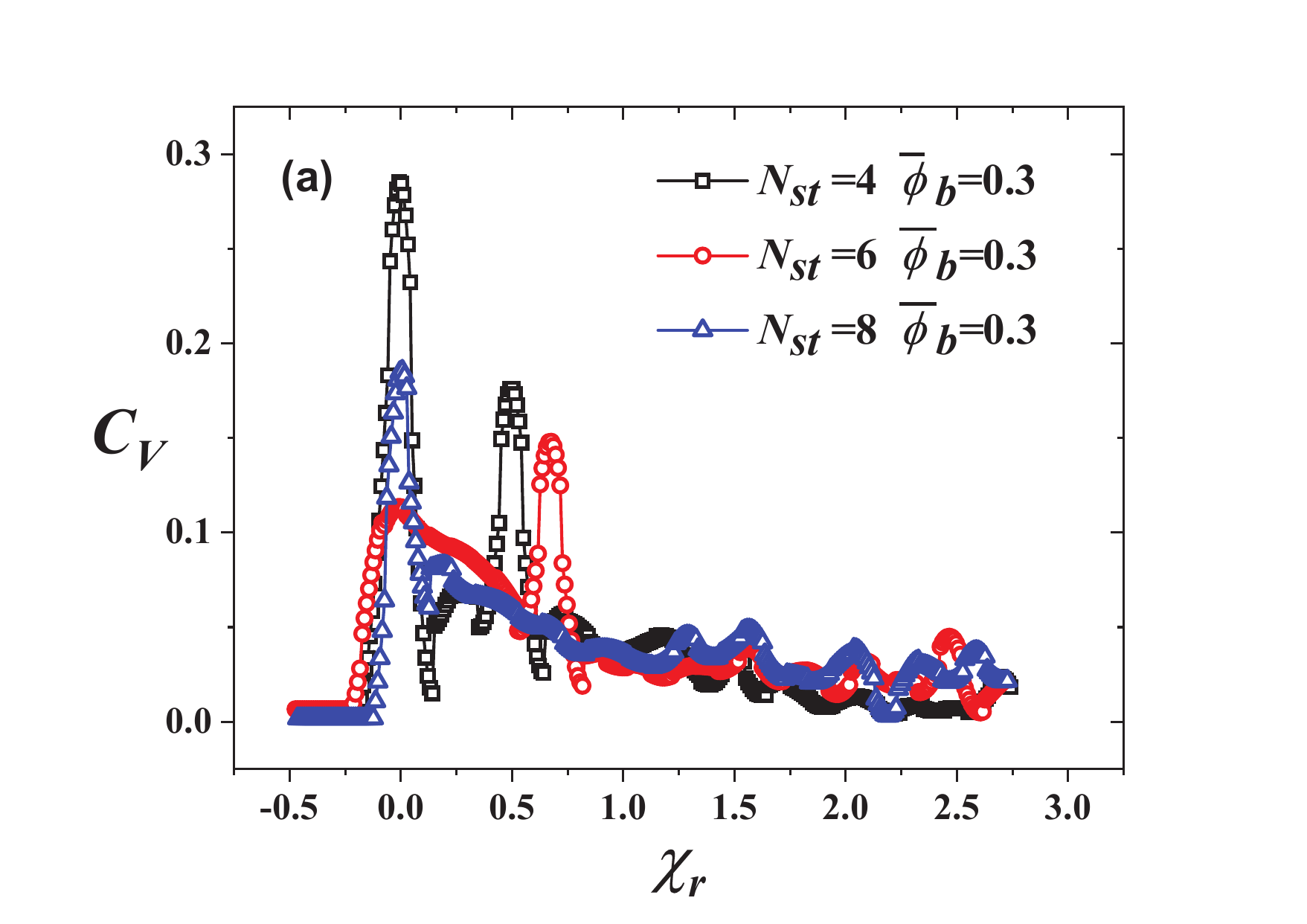}
\includegraphics[width=5cm]{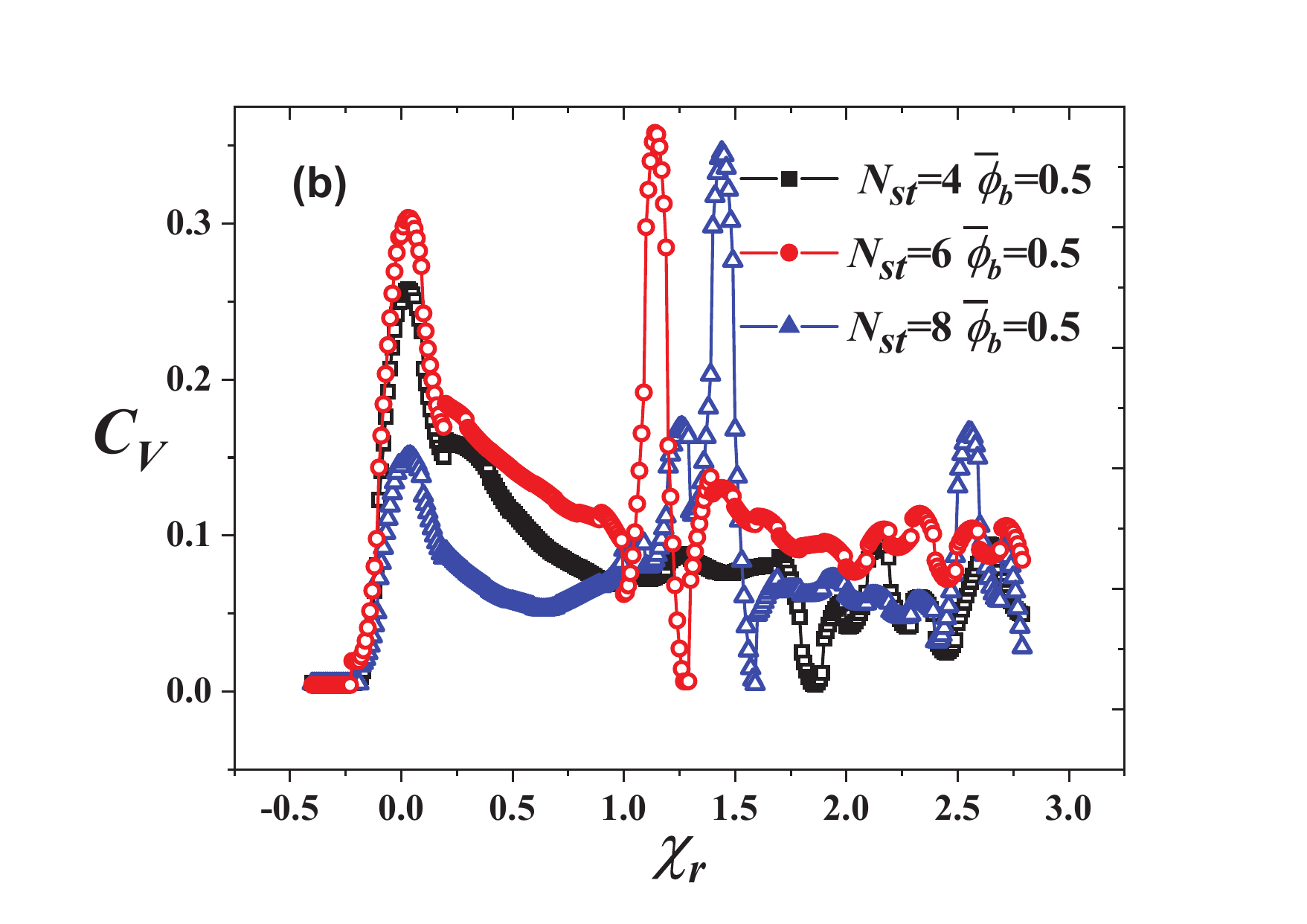}
\includegraphics[width=5cm]{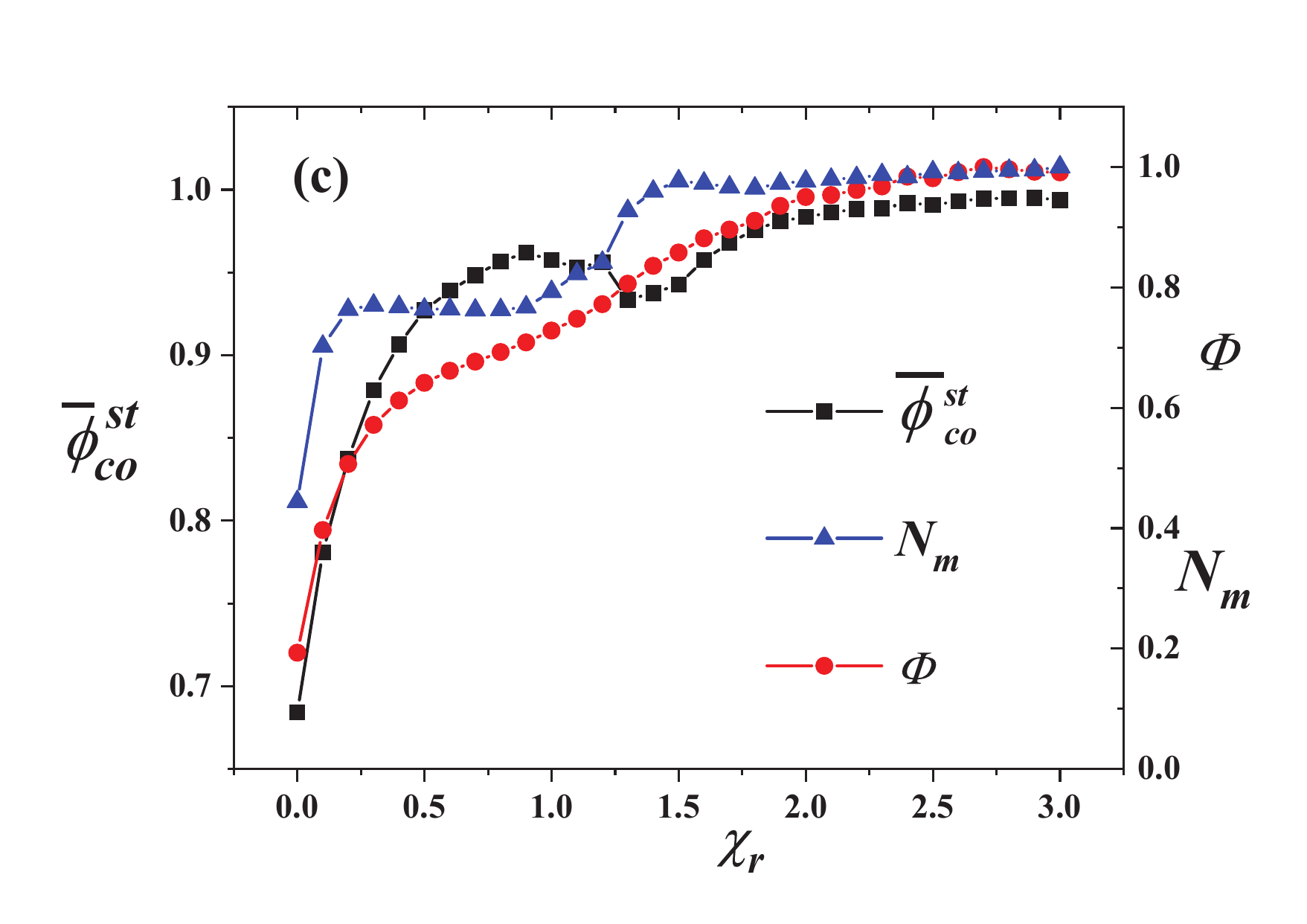}
\includegraphics[width=5cm]{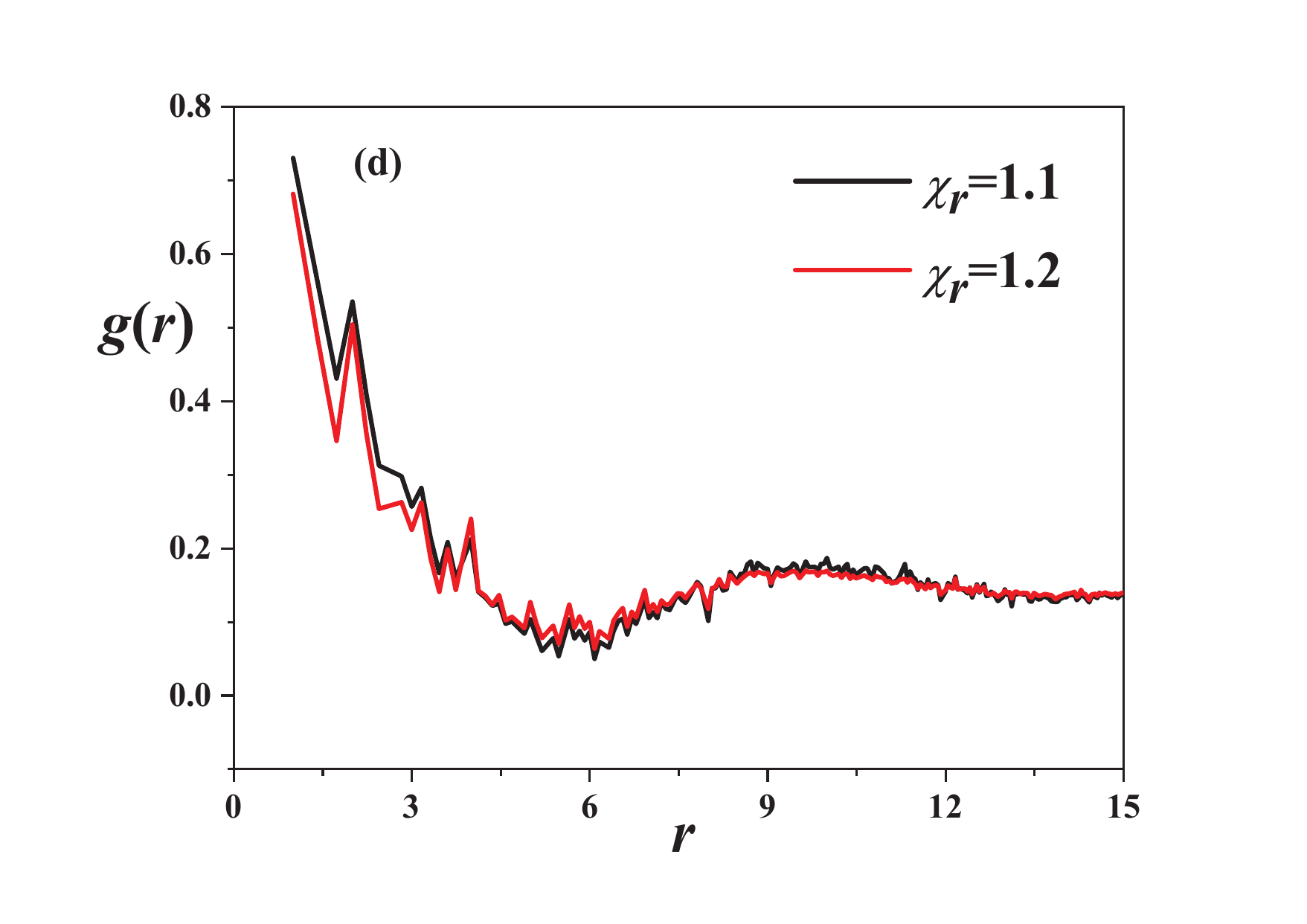}
\caption{(Colour online) Equilibrium aggregation curve and the radical distribution function: (a) for $\overline{{{\phi }}}_{b}=0.3$; (b), (c), (d) for $\overline{{{\phi }}}_{b}=0.5$. (c), (d) only for ${{N}_{\text{st}}}= 8$, $6$, respectively.}
\label{fig6}
\end{figure}

At intermediate concentrations ($\overline{\phi }_{b}=0.3$, figure~\ref{fig6}a), two peaks appear at ${{N}_{\text{st}}}=4$. The first is related to the appearance and growth of lamellar and rod micelles. The second involves micelle fusion, where some lamellar/rod micelles transform into cubic micelles, accompanied by a considerable increase in micellar core volume and a decrease in the average aggregation degree (not shown). As the rod block length increases, the height and number of heat capacity peaks decrease sequentially. The shape of cubic micelles with interaction parameter does not change. Accordingly, the fusion between micelles becomes easier, and the growth of micelles becomes smoother, that is, the rearrangement behavior is weakened. Compared to the solutions~\cite{Han2022}, homopolymer addition promotes micelle fusion, which is similar to the effect of the increase in concentration on the aggregation behavior.
\begin{figure}[!t]
\centering
\includegraphics[width=14cm]{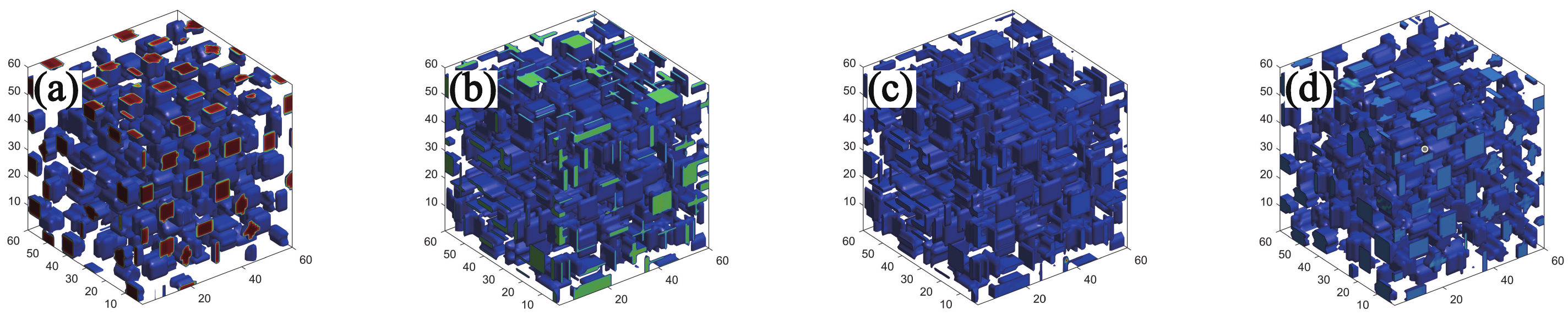}
\caption{(Colour online) The  micellar structures  for $\overline{{{\phi }}}_{b}=0.5$, ${{N}_{\text{st}}}=8$ (${{\phi }_{A}}>0.5$). (a) ${{\chi }_{r}}=1.3$, (b)~${{\chi }_{r}}=1.5$, (c) ${{\chi }_{r}}=2.5$, (d) ${{\chi }_{r}}=2.6$.
\label{fig7}
}
\end{figure}

At relatively high concentrations ($\overline{{\phi }}_{b}=0.5$, figure~\ref{fig6}b), the dependence of rearrangement behavior on ${{N}_{\text{st}}}$
strengthens. Unlike in low and intermediate concentrations, the increase in the rod blocks promotes the rearrangement. At ${{N}_{\text{st}}}=4$, only one heat capacity peak appears, related to the coexistence of  cubic and lamellar micelles and their fusion. At ${{N}_{A}}=6$ (figure~\ref{fig7}c), two peak appear. The first peakas shown corresponds to the transformation of spherical to cubic micelles and the gradual appearance of lamellar micelles. The second peak, especially at ${{\chi }_{r}}=1.2$, is caused by a sharp increase in the lamellar fraction of micelles, i.e., the micellar laminarization. To quantify this laminarization, the radical distribution function of the volume faction of stickers at the micellar cores, $g(r)=\langle\phi^{co} _{st}(r)\phi^{co} _{st}(0)\rangle$, is calculated as shown in (figure~\ref{fig6}d), which verifies the micellar shape change. The increase in rod block length accelerates the effect of interaction parameters on the structure. At ${{N}_{\text{st}}}=8$, three  peaks appear. The first peak is related to the appearance of ordered spherical and cubic micelles. The second peak, at ${{\chi }_{r}}=1.5$ is directly related to micellar laminarization (figure~\ref{fig7}a,b). Notably, with a further parameter increase, inverse laminarization corresponding to the third peak occurs, marked by a remarkable increase in cubic micelles at ${{\chi }_{r}}=2.6$ (figure~\ref{fig7}c,d). Furthermore, longer rod blocks accelerate the cooperative variation of 
$\overline{{{\phi }}}_{co}^{st}$ and $N_m$ with ${{\chi }_{r}}$
observed between $\chi_{r}$ between ${{\chi }_{r}}=0.5$ and ${{\chi }_{r}}=2.0$ (figure~\ref{fig6}c). Homopolymer adding  attributes to both cooperative growth which emerges at higher concentrations and  the arrangement related to inverse laminarization. 
\begin{figure}[!t]
\centering  
\includegraphics[width=6cm]{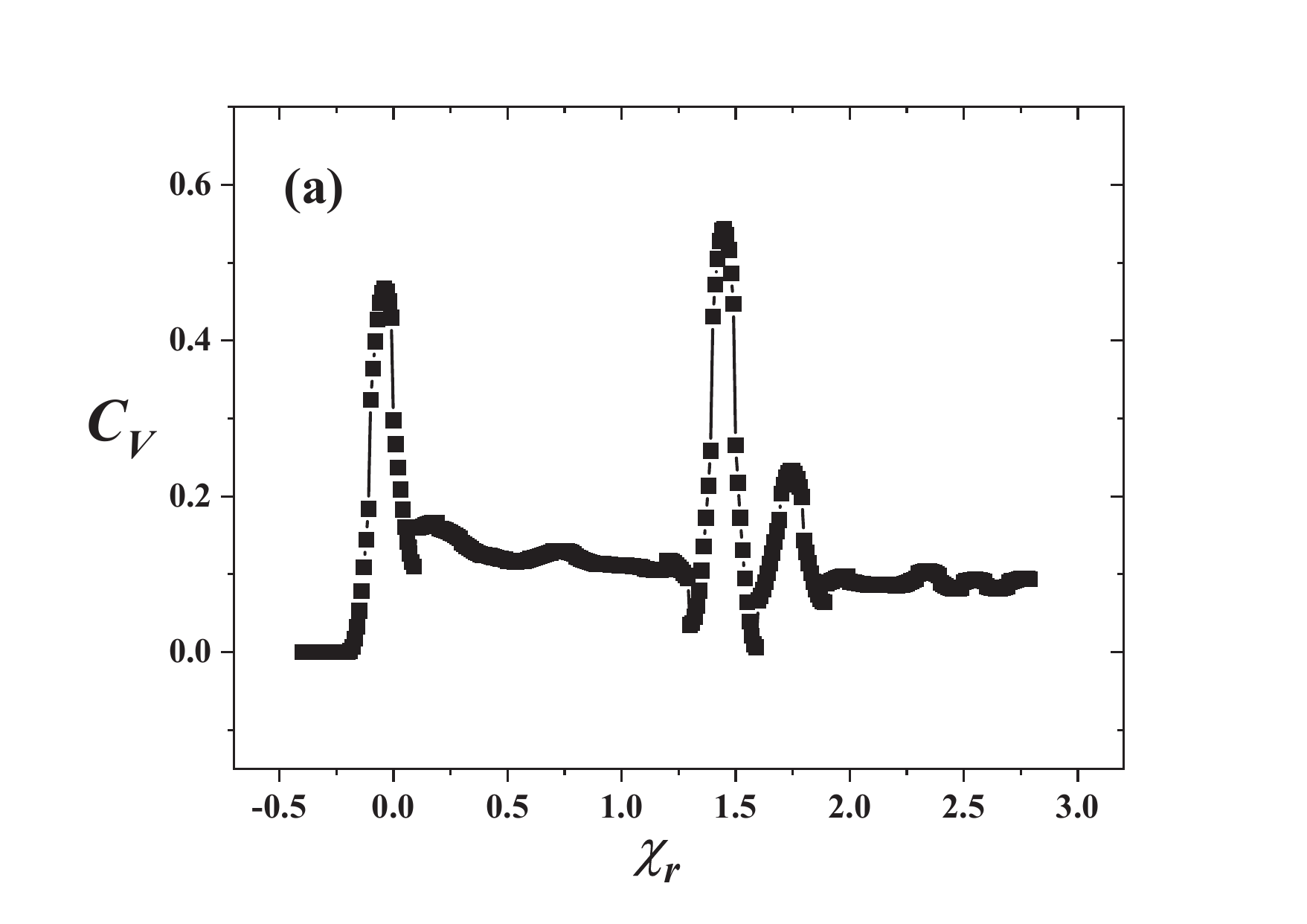}
\includegraphics[width=6cm]{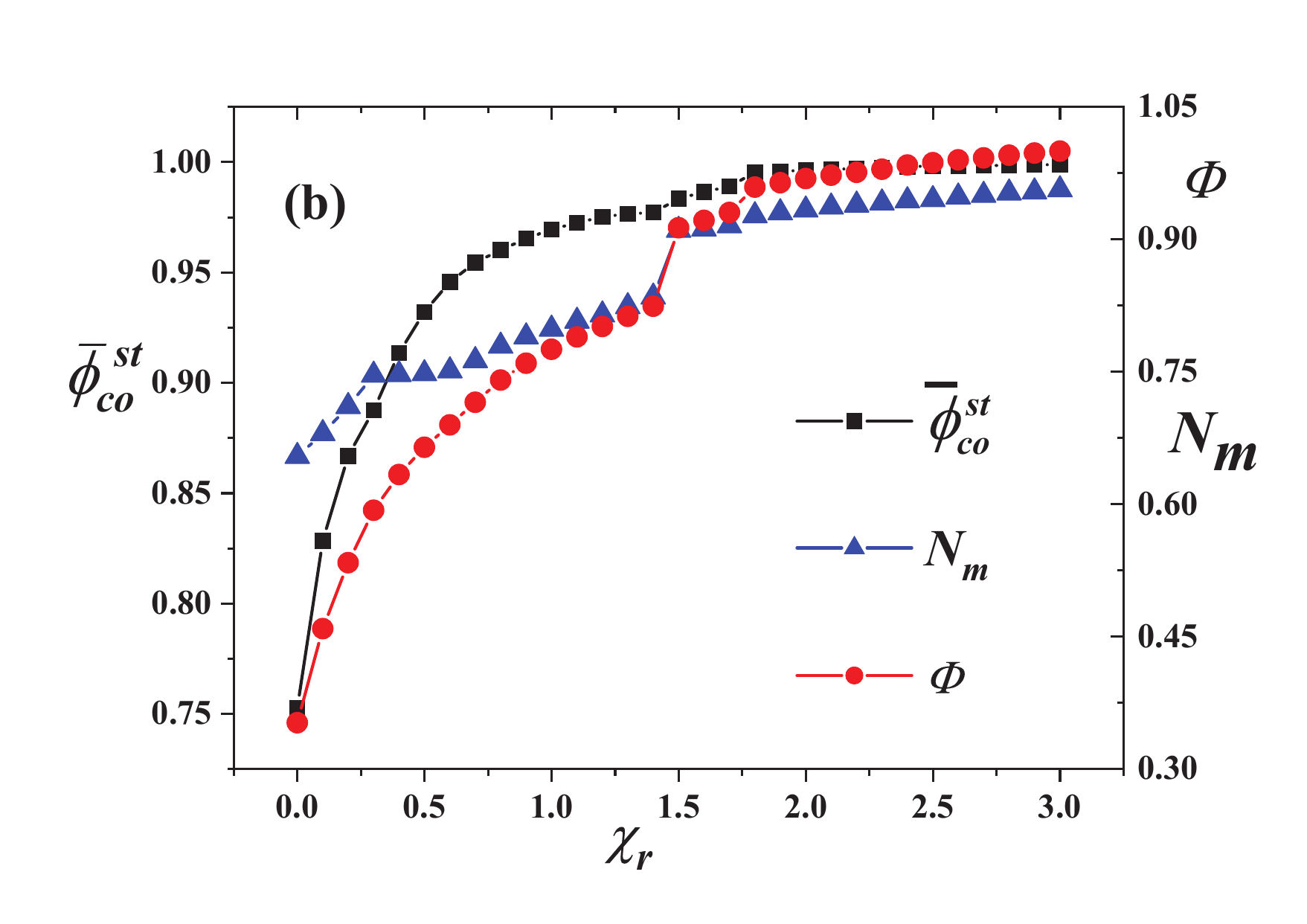}
\caption{(Colour online) Equilibrium aggregation curve of $\overline{{{\phi }}}_{b}=0.8$, ${{N}_{\text{st}}}=4$.}
\label{fig8}
\end{figure}
In the solutions ($\overline{{\phi }}_{b}=0.8$)~\cite{Han2022}, the emergence of the third peak is related to the ordering of lamellar micelle. It should result from the reduction of interfacial tension. In this work, the length of homopolymer is smaller than that of coil blocks. { Homopolymer chains permeate and swell the coil block, which preferentially is in the direction parallel to the lamellar interface, resulting in the the wet-brush appearance~\cite{Hayashi2019}. This leads to the crowding effect compared with that of the solvents. Accordingly, a remarked reduction of the configurational entropy of the coil block leads to the inverse laminarization. Similarly the cylindrical /spherical phase emerges in coil/rod/coil BAB triblock copolymer melts with an increase of the Flory-Hugins interaction~\cite{Chen2007}. By constrast, when the length of homopolymers is larger than that of a coil block.} The homopolymers segregate among the micelles, where the dry-brush tends to form~\cite{Hayashi2019}. Consequently, the  strengthening depletion is favorable to the increase in the aggregation extent. However, the inverse laminarization will not occur probably. 
\begin{figure}[!t]
\centering
\includegraphics[width=14cm]{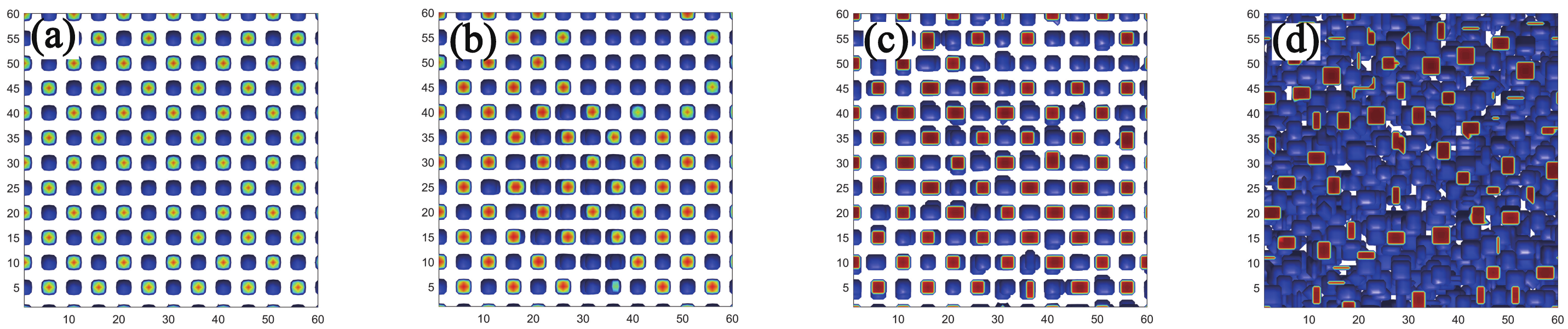}
\caption{(Colour online) The micellar structures for $\overline{{{\phi }}}_{b}=0.8$, ${{N}_{\text{st}}}=4$ (${{\phi }_{st}}>0.5$). (a) ${{\chi }_{r}}=0.0$, (b)~${{\chi }_{r}}=0.1$, (c) ${{\chi }_{r}}=1.4$, (d) ${{\chi }_{r}}=1.5$.
\label{fig9}
}
\end{figure}
\begin{figure}[!t]
\centering  
\includegraphics[width=5cm]{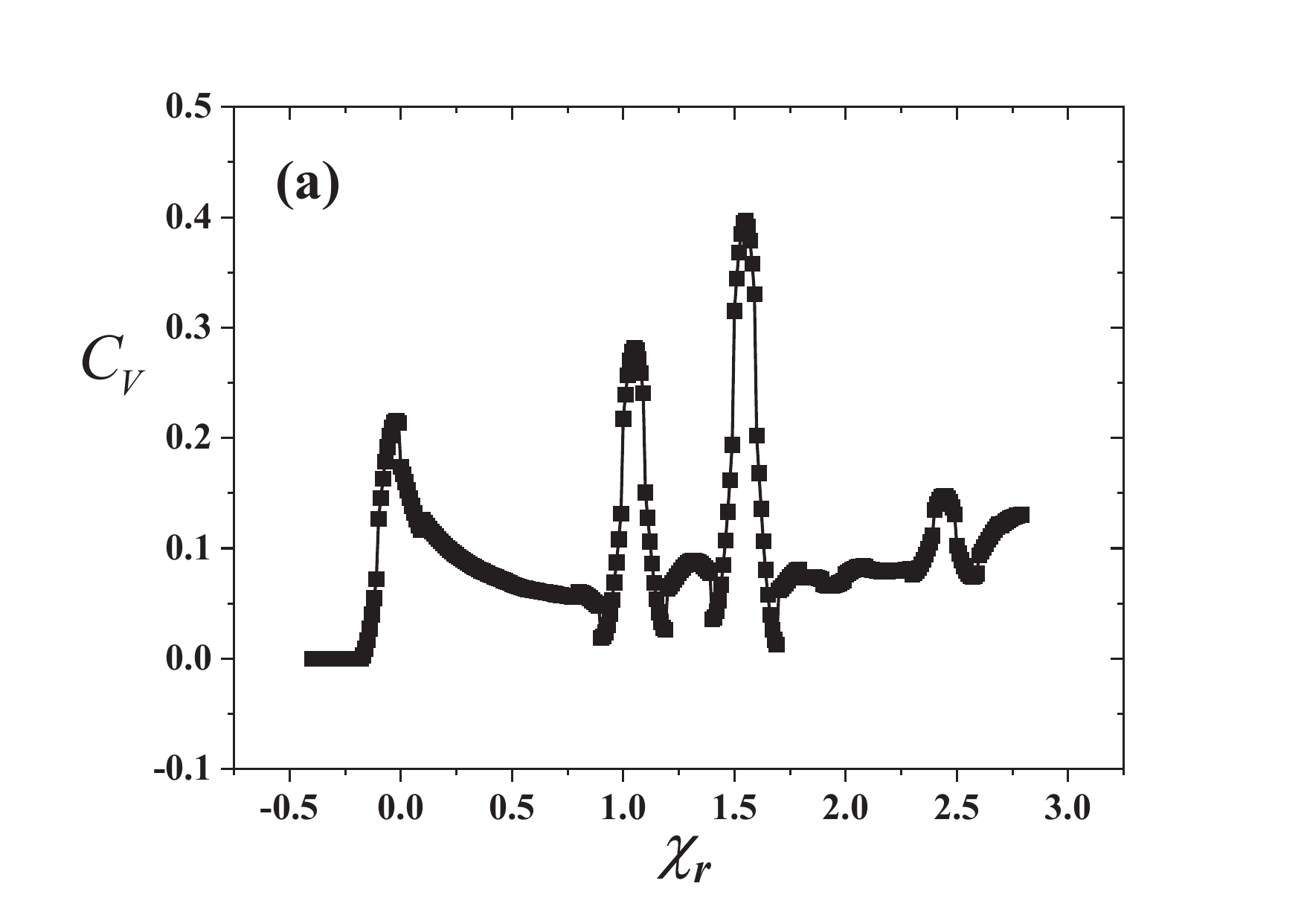}
\includegraphics[width=5cm]{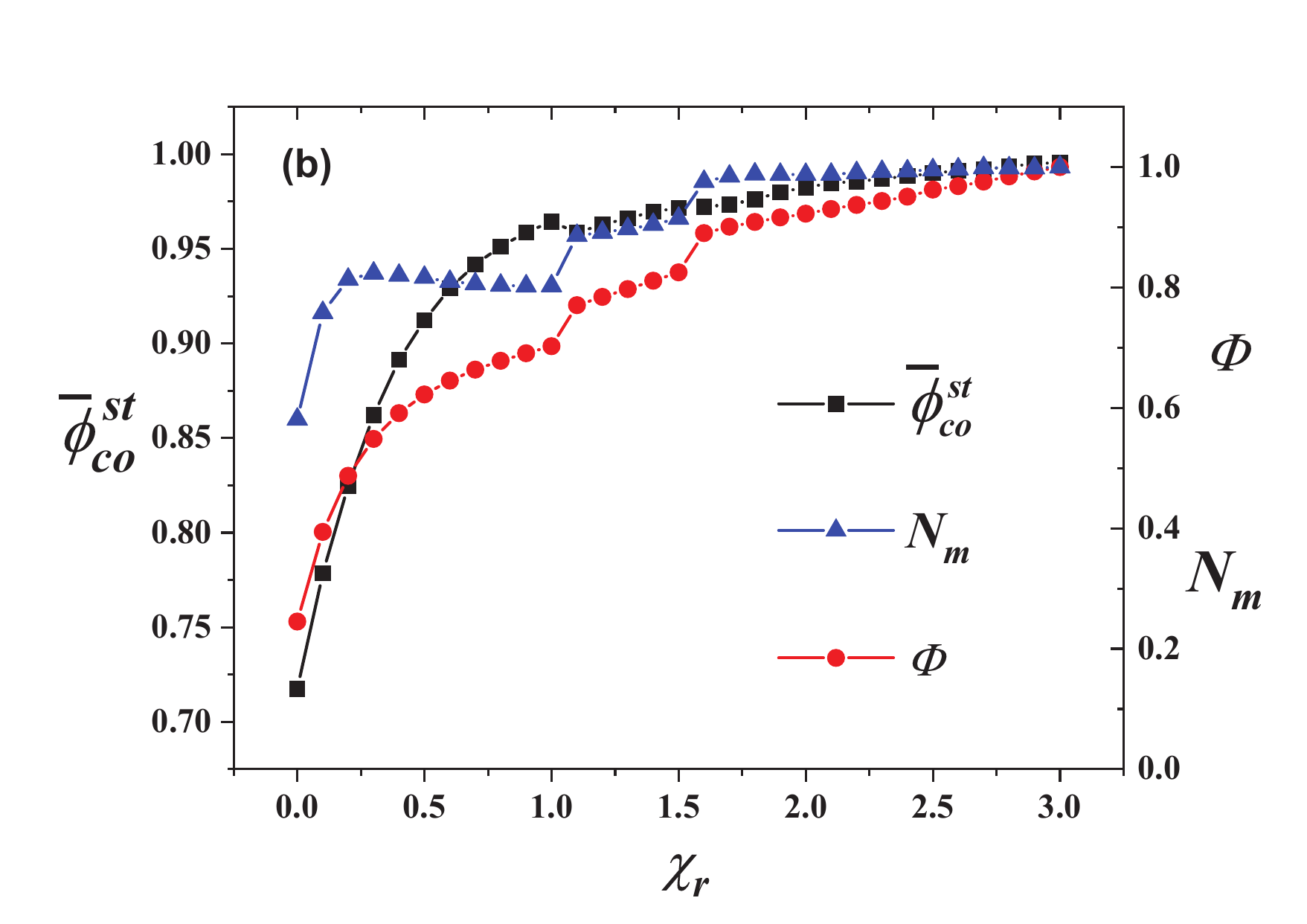}
\includegraphics[width=5cm]{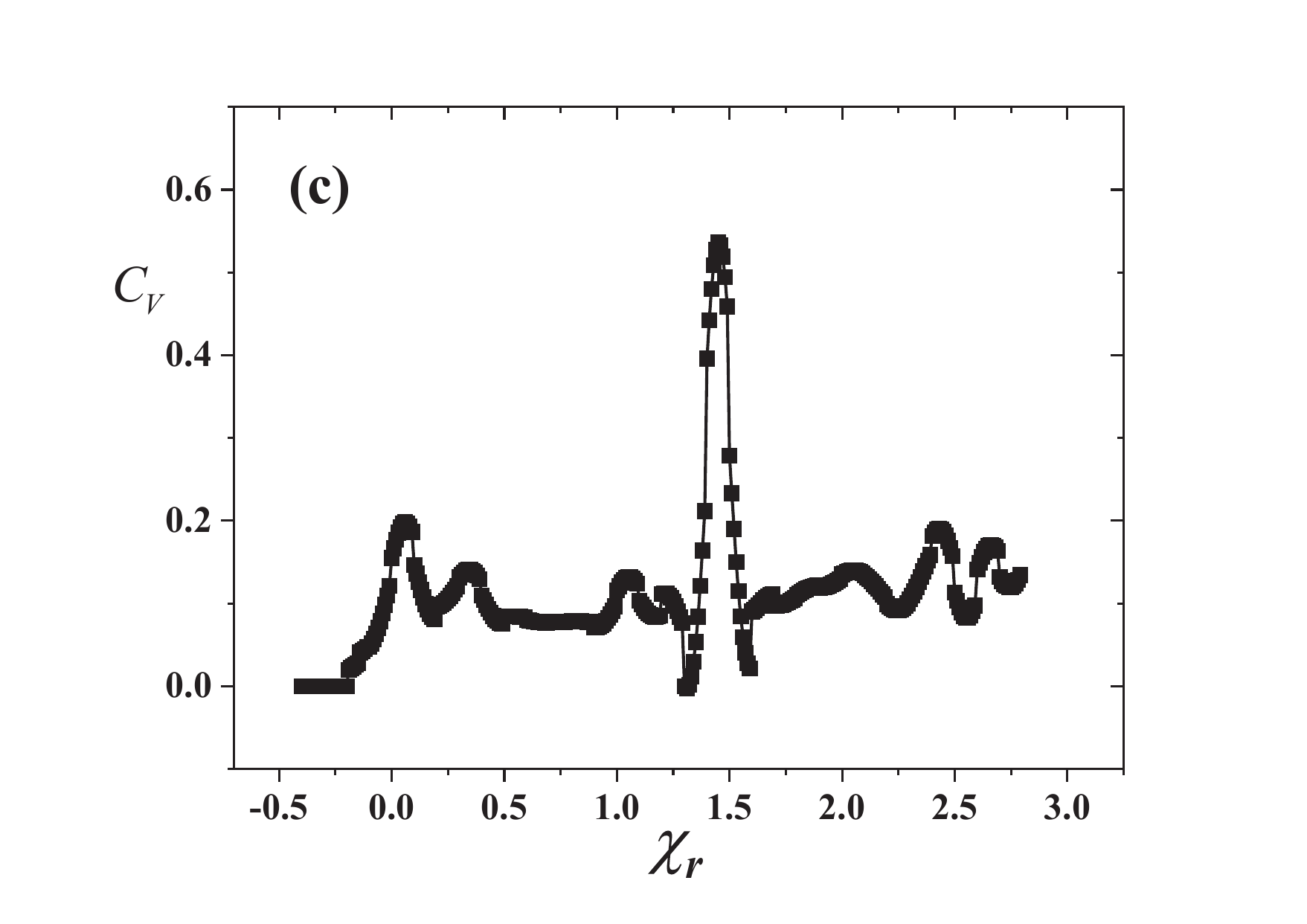}
\includegraphics[width=5cm]{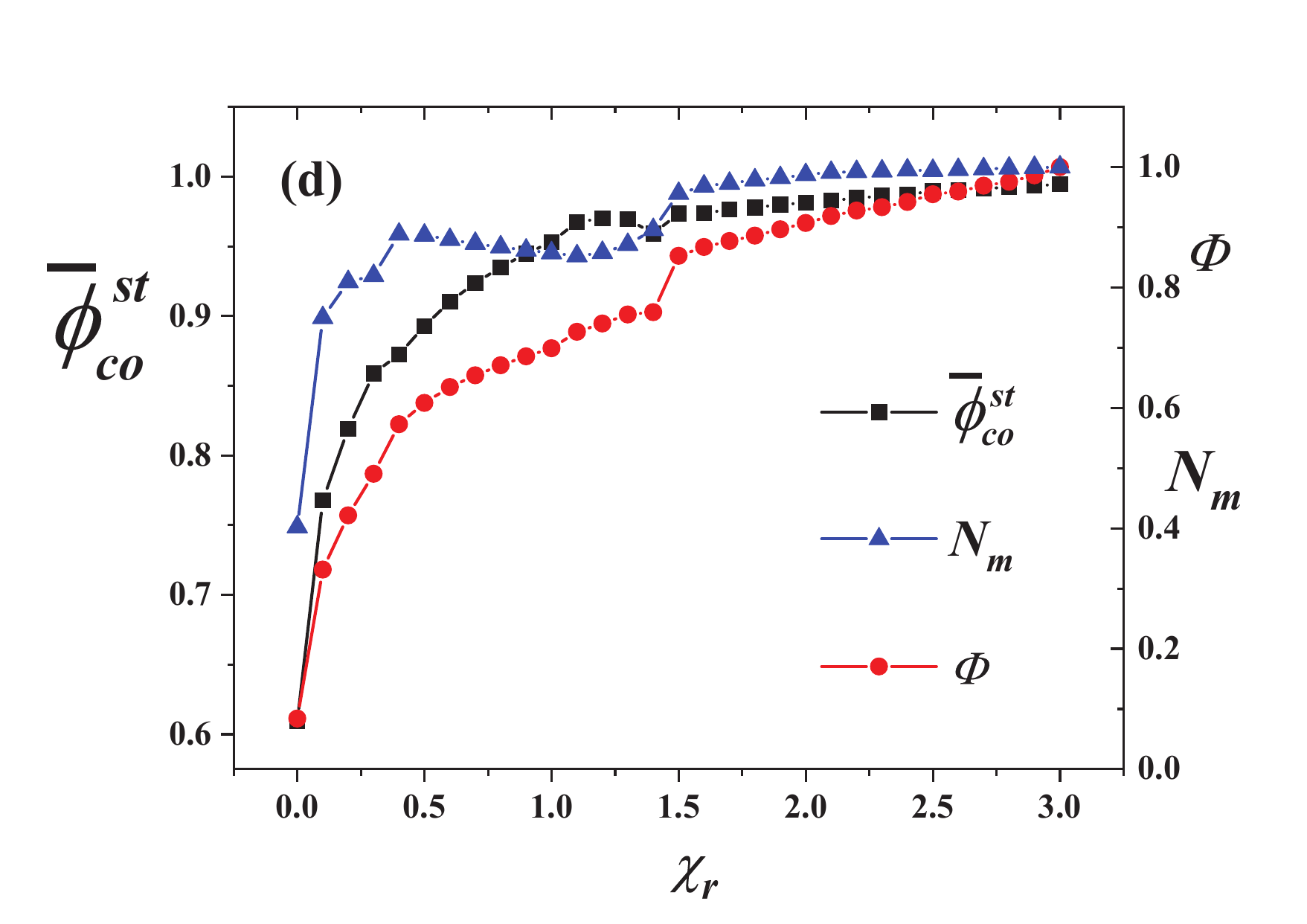}
\includegraphics[width=5cm]{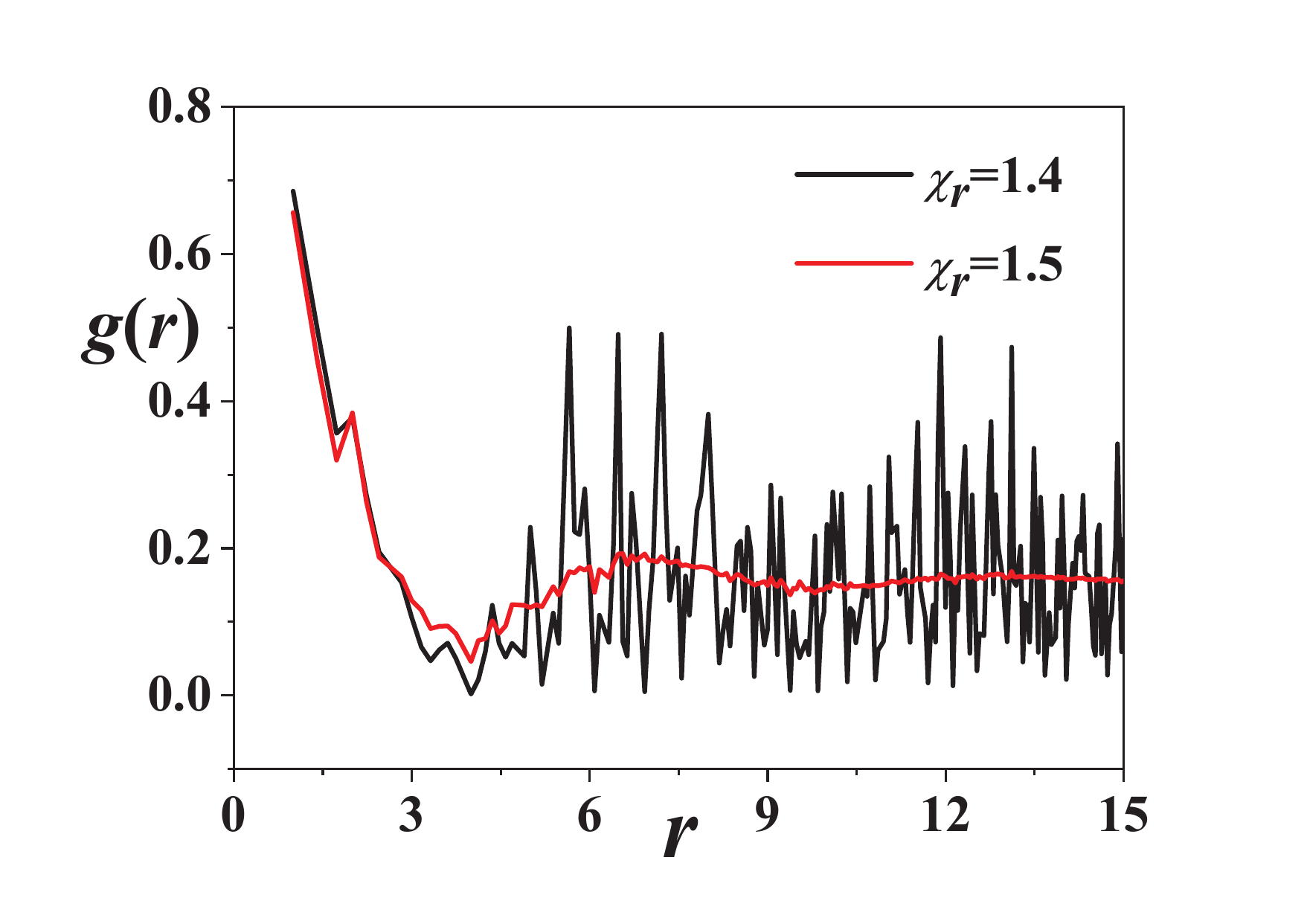}
\includegraphics[width=5cm]{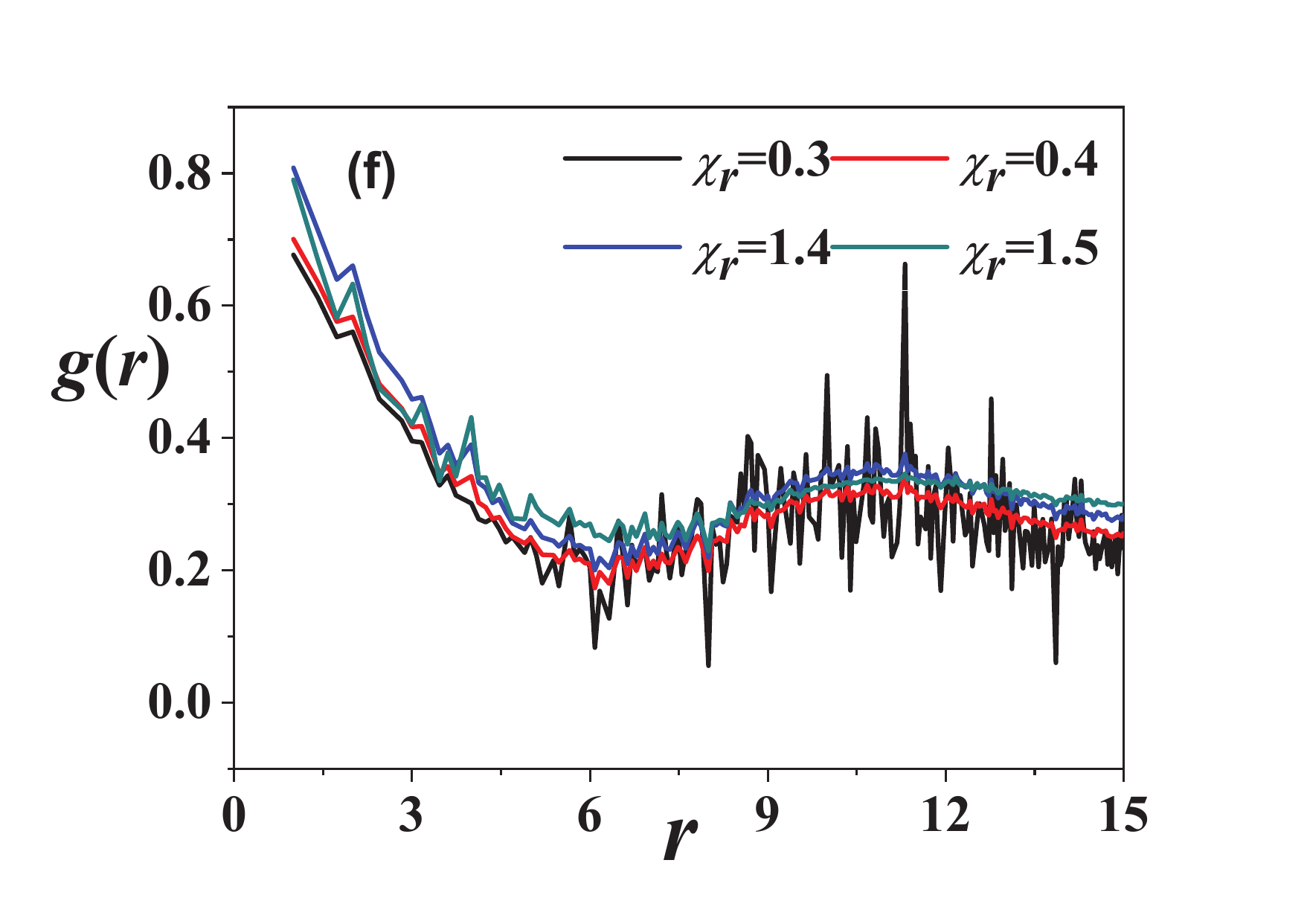}
\caption{(Colour online) Equilibrium aggregation curve and radical distribution function of $\overline{{{\phi}}}_{b}=0.8$ for different $N_{\text{st}}$: (e) for  ${N}_{\text{st}}=4$; (a), (b)~for  ${N}_{\text{st}}=6$; (c), (d) and (f) is related to  ${N}_{\text{st}}=8$.}
\label{fig10}
\end{figure}

\begin{figure}[!t]
\centering
\includegraphics[width=14cm]{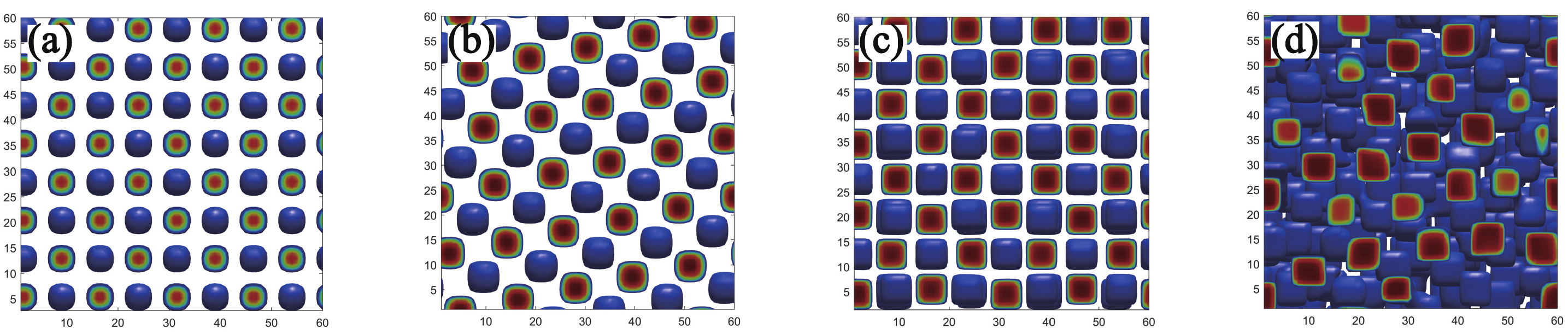}
\caption{(Colour online) The  micellar structures  for $\overline{{{\phi }}}_{b}=0.8$, $N_{\text{st}}=8$ (${{\phi }_{st}}>0.5$). (a) ${{\chi }_{r}}=0.0$, (b)~${{\chi }_{r}}=0.1$, (c)~${{\chi }_{r}}=0.2$, (d) ${{\chi }_{r}}=0.4$.
\label{fig11}
}
\end{figure}
At high concentrations ($\overline{{{\phi }}}_{b}=0.8$), the addition of homopolymers accelerates rearrangement for small ${{N}_{\text{st}}}$, but suppresses for large ${{N}_{\text{st}}}$. At ${{N}_{\text{st}}}=4$ (figure~\ref{fig8}), three heat capacity peaks appear. The first corresponds to the appearance of ordered cubic micelles (figure~\ref{fig9}a), which become irregular, yet remain overall ordered as ${{\chi }_{r}}$
increases (figure~\ref{fig9}b). The second peak, especially at ${{\chi }_{r}}=1.5$, involves a transition to disordered micelles, with more irregular cubic shapes and the appearance of lamellar micelles (figure~\ref{fig9}c,d), confirmed by a change of the radial distribution function (figure~\ref{fig10}e). The third peak shows no obvious morphological change but is linked to simultaneous increases in the average aggregation degree and micellar core volume due to fragmentation/fusion. In the solutions~\cite{Han2022}, the system does not exhibit an order/disorder transition which occurs at ${{N}_{\text{st}}}=4$, but instead at ${{N}_{\text{st}}}=6$. It is demonstrated that the effect of adding a homopolymer is also similar to the increase in rod block length. At intermediate rod length (${{N}_{\text{st}}}=6$), four peaks appear (figure~\ref{fig9}a). The first two peaks correspond to the emergence of ordered spherical micelles and an order/disorder transition with laminarization, respectively. The third and fourth peaks lack obvious morphology changes and are probably related to fragmentation/fusion, primarily increasing the micellar core volume. Although the number of heat capacity peaks increases and the rearrangement behavior increases, their corresponding structural change characteristics remain basically unchanged.

With the further increase in rod block length, the characteristics of the system ordering transformation become more abundant. At $N_{\text{st}}=8$, three peaks appear (figure~\ref{fig10}c). The first peak has two sub-peaks: the first relates to a ``reversible'' change in micelle packing from aligned perpendicular to the coordinate axes, to non-perpendicular, and back, accompanied by spherical/cubic transitions (figure~\ref{fig11}a--c). The second sub-peak is due to the transformation of ordered micelles into disordered, irregular ones (figure~\ref{fig11}d). The second peak is caused by a laminarization of cubic micelles, confirmed by a sharp drop in the first minimum of the radial distribution function at ${{\chi }_{r}}=1.5$ (not shown), accompanied by fragmentation/fusion. The third peak is also associated with fragmentation/fusion, and the micellar structures almost remain unchanged. The inverse laminarization does not occur, unlike the $N_{\text{st}}=8$ case at $\overline{{{\phi }}}_{b}=0.5$. Under the condition of $\overline{{{\phi }}}_{b}=0.8$, the effect of configurational entropy reduces, and the effect of interfacial interaction rises. At high concentrations, the effect of  adding homopolymer is weakened for a larger rod monomer concentration, whose tendency is similar to the experimental results~\cite{Hayashi2019}. Compared to the solutions, homopolymer addition is advantageous to micelle rearrangement at $N_{\text{st}}=4$. { For large $N_{\text{st}}$ ($=6,8$) is suppressed, especially at large ${{\chi }_{r}}$.} The disorder/order transition at $N_{\text{st}}=6$ disappears, and the cooperative growth  at $N_{\text{st}}=8$ is suppressed.

\section{Conclusion and summary} \label{sec4}

Using the lattice self-consistent field theory,  the effect of the addition of B homopolymer on the aggregation behavior was investigated in amphiphilic coil/rod/coil BAB triblock copolymers. It depends on the length of the hydrophobic rod block and copolymer concentration which shows a non-monotonous  concentration-dependent effect. At low concentrations, homopolymer addition suppresses the rearrangement behaviors concerned with the appearing times of micellar diffusion and rod/lamella transitions for short rod blocks, while it enriches the structural behavior, such as the emergences of cubic and large lamellar micelles, and enriches the rearrangement mechanism inducing orientation-dependent diffusion and inverse diffusion processes for long rod blocks. At intermediate concentrations, the primary consequence is the promotion of micelle fusion, and smoothing the growth process. This indicates that homopolymer accelerates the effect of increasing concentration. At relatively high concentrations, unlike in low and intermediate concentrations, both the addition of homopolymers and the increase in rod block length promote rearrangement. For long rod block case,  adding homopolymers leads to the arrangement related to inverse laminarization. { The cooperative growth was found to emerge at high concentration of the solutions.} At high concentrations, the impact of  addition of homopolymer weakens as the rod block elongates. The effect of the addition of homopolymer depends on rod blocks. Compared to the solutions, for a short rod block system, homopolymer addition is advantageous to micelle rearrangement concerned with order/disorder transition, and  the rearranement in the system of intermediate length and long rod block  is suppressed, especially to large ${{\chi }_{r}}$. The disorder/order transition at $N_{\text{st}}=6$ disappears, and the cooperative growth  at $N_{\text{st}}=8$ is suppressed. 
 
 In essence, the addition of homopolymer introduces a competing effect. It constrains the configurational entropy regulation of the coil blocks, thereby modulating the capability of the system to undergo structural transformations driven by the maximization of interfacial energy. { This leads to a complex phase behavior, where the most pronounced structure peculiarities and orientation change.} This work provides theoretical insights into  the self-assembly in complex block copolymer blends containing rigid blocks, offering  guidance for the experimental design of functional nanomaterials with targeted nanostructures.

\section*{Acknowledgments}
This work is subsidized by the National Natural Science Foundation
of China Programs (22163005).

\ukrainianpart
\title
{Вплив додавання гомополімерів на агрегаційну поведінку в амфіфільних триблокових кополімерах, що містять жорсткі блоки %
}
\author
{С.-Г. Хан\orcid{0000-0002-2724-0114}, Х. Жанг\orcid{0009-0006-2892-8021}, З. Х. Сун, M.~ І.~Жу }
\address{
	Науково-технічний факультет Внутрішньомонгольського університету науки і технологій, Баотоу 014010, Китай}

\makeukrtitle
\begin{abstract}
	Вплив додавання гомополімеру B на агрегаційну поведінку амфіфільних триблокових кополімерів BAB типу клубок/стрижень/клубок було досліджено за допомогою ґраткової теорії самоузгодженого поля. Цей вплив залежить від довжини гідрофобного стрижневого блоку та концентрації кополімеру. Порівняно з розчинами, при низьких концентраціях додавання гомополімеру сприяє появі кубічних та великих пластинчастих міцел. Хоча це і збагачує структурну поведінку, перегрупування пригнічується. Зі збільшенням стрижневих блоків перегрупування пов'язане з процесами дифузії, що залежить від орієнтації, та зворотної дифузії. При відносно високих концентраціях додавання гомополімерів сприяє перегрупуванню. У випадку блоків з довгими стрижнями виникає перебудова, пов’язана зі зворотною ламінаризацією та кооперативним зростанням. При високих концентраціях для системи з короткими стрижнями додавання гомополімерів є вигідним для перебудови міцел, пов'язаної з переходом порядок/безлад, а перебудова в системі з середніми та довгими стрижнями пригнічуються. Ця робота допомагає зрозуміти механізм росту міцел з жорсткими блочними ядрами.
	\keywords перегрупування, розчинення міцел, блок-кополімер клубок/стрижень/клубок, теорія самоузгодженого поля
\end{abstract}

  \lastpage
 \end{document}